\documentclass[11pt]{article}

\usepackage{scicite}

\usepackage{times}

\usepackage[dvips]{graphicx}

\newcommand{\bvec}[1]{\mbox{\boldmath $#1$}}

\def\aj{{Astron. J.}}
\def\araa{{Ann. Rev. Astr. Astrophys.}}
\def\apj{{Astrophys. J.}}
\def\apjs{{Astrophys. J. Suppl.}}
\def\apss{{Astrophys. Space Sci.}}
\def\mnras{{Mon. Not. R. Astr. Soc.}}
\def\nat{{Nature}}
\def\pasp{{Publ. Astron. Soc. Pacific}}
\def\apjl{{Astrophys. J.}}

\def\prd{{Phys. Rev. D}}
\def\aap{{Astr. Astrophys.}}
\def\VAR#1{{\rm Var}( #1)}
\def\ell{l}
\newenvironment{sciabstract}{%
\begin{quote} \bf}
{\end{quote}}

\newcounter{lastnote}

\title{Polarization Observations with the Cosmic Background Imager}

\author{A. C. S. Readhead,$^{1\ast}$
\and S. T. Myers,$^{2}$
\and T. J. Pearson,$^{1}$
\and J. L. Sievers,$^{1,3}$
\and B. S. Mason,$^{4}$
\and C. R. Contaldi,$^{3}$
\and J. R. Bond,$^{3}$
\and R. Bustos,$^{5}$
\and P. Altamirano,$^{6}$
\and C. Achermann,$^{6}$
\and L. Bronfman,$^{6}$
\and J. E. Carlstrom,$^{7}$
\and J. K. Cartwright,$^{1,7}$
\and S. Casassus,$^{6}$
\and C. Dickinson,$^{1}$
\and W. L. Holzapfel,$^{8}$
\and J. M. Kovac,$^{1,7}$
\and E. M. Leitch,$^{7}$
\and J. May,$^{6}$
\and S. Padin,$^{1,7}$
\and D. Pogosyan,$^{3,9}$
\and M. Pospieszalski,$^{10}$
\and C. Pryke,$^{7}$
\and R. Reeves,$^{5}$
\and M. C. Shepherd,$^{1}$
\and S. Torres$^{5}$    
\\ 
\normalsize{$^{1}$Owens Valley Radio Observatory, California Institute of Technology, Pasadena, CA 91125, USA.}\\
\normalsize{$^{2}$National Radio Astronomy Observatory (NRAO), Post Office Box O, Socorro, NM 87801, USA.}\\
\normalsize{$^{3}$Canadian Institute for Theoretical Astrophysics (CITA), University of Toronto, }\\
\normalsize{Toronto, Ontario, M5S3H8, Canada.}\\
\normalsize{$^{4}$NRAO, Post Office Box 2, Green Bank, WV 24944, USA}\\
\normalsize{$^{5}$Departamento de Ingenier{\'\i}a El{\'e}ctrica, Universidad de Concepci{\'o}n, Concepci{\'o}n, Chile.}\\
\normalsize{$^{6}$Departamento de Astronom{\'\i}a, Universidad de Chile,  Santiago, Chile.}\\
\normalsize{$^{7}$Kavli Institute for Cosmological Physics, Department of Astronomy and Astrophysics,}\\
\normalsize{University of Chicago, Chicago, IL 60637, USA.}\\
\normalsize{$^{8}$Department of Physics, 361 LeConte Hall, University of California, Berkeley, CA 94720--7300.}\\
\normalsize{$^{9}$Department of Physics, University of Alberta, Edmonton, Alberta,  T6G2J1, Canada.}\\
\normalsize{$^{10}$NRAO, 520 Edgemont Road, Charlottesville, VA 22903, USA.}\\
\normalsize{$^\ast$To whom correspondence should be addressed; E-mail:  acr@astro.caltech.edu.}\\\\
\normalsize{\bf Science 306: 836-844. Published online October 7, 2004; 10.1126/science.1105598}
}

\date{}

\begin{document} 

\baselineskip14pt

\maketitle 

\begin{sciabstract}
  
  Polarization observations of the cosmic microwave background with
  the Cosmic Background Imager from September 2002 to May
  2004 provide a significant detection of the E-mode
  polarization and reveal an angular power spectrum of
  polarized emission showing peaks and valleys that are shifted in
  phase by half a cycle relative to those of the total intensity
  spectrum.  This key agreement between the
  phase of the observed polarization spectrum and that predicted
  on the basis of the total intensity spectrum provides support for the
  standard model of cosmology, in which dark matter and dark energy
  are the dominant constituents, the geometry is close to flat,
  and primordial density fluctuations are predominantly adiabatic with
  a matter power spectrum commensurate with inflationary cosmological
  models.

\end{sciabstract}


\noindent In recent years a wide variety of observations have provided 
support for a standard model of cosmology and cosmic structure
formation.  In this model \cite{Spergel03,Riess04,Tegmark04}, 
the mass-energy density of the universe
is dominated by cold dark matter and dark energy, possibly in the form
of Einstein's cosmological constant, and conventional baryonic matter is
only a minor component.  As the universe expanded from its hot, dense
origins, all the structures seen in the universe today formed under
the action of gravity on initial nearly scale-invariant adiabatic
gaussian density fluctuations.  Observations of the anisotropy in the intensity of
the cosmic microwave background radiation (CMB) have provided much of
the evidence for this model and estimates of the values of the
fundamental parameters, including the spatial curvature and the
densities of dark energy, cold dark matter, and ordinary matter. The
two major ingredients of the standard model, dark matter and dark
energy, are far from understood, and their existence presents the
most serious challenge to physics since the quantum and relativistic
revolutions of a century ago. It is therefore essential to extract as
much information as possible from the observations in order to test
all aspects of the standard model and to look for possible anomalies
that might provide insights into the nature of these two dark
components of the cosmos.

In the 1980s successively more stringent limits were placed on the
observed level of temperature anisotropy in the CMB
\cite{Readhead89,Readhead92}, providing convincing evidence that the
dominant matter constituent in the universe is non-baryonic.  These
searches culminated in the detection of anisotropies by the COBE
satellite \cite{Bennett94}, a confirmation of one of the major
predictions of theoretical cosmology
\cite{Peebles70,Sunyaev70,Bond87,1978AZh....55..913D,1981ApJ...243...14W,1984ApJ...285L..39V}.
Rapid advances in experimental techniques have since delineated the
prominent features in the angular power spectrum
\cite{Netterfield97,Church97,Miller99,Leitch00,debernardis00,Hanany00}.
The spectrum of fluctuations on large angular scales [low multipole
numbers, $l < 500$ \cite{multipole}] has been measured with high
precision by the {\it Wilkinson Microwave Anisotropy Probe} (WMAP)
\cite{Hinshaw03}, whereas precise and sensitive CMB observations from
the ground and from balloon-based platforms have extended the spectrum to
angular scales as small as a few minutes of arc ($l\approx 3500$)
\cite{Halverson02,Benoit03,Dickinson04,Readhead04,Kuo04}.

The early universe was opaque to electromagnetic radiation, but as it
expanded and cooled the hot electron-baryon plasma combined into
neutral hydrogen and helium and the universe became transparent. The
microwave background photons that we detect today have passed freely
through the universe since they were last scattered by electrons in
the ionized plasma. They thus provide a picture of the physical
conditions at the time of last scattering when the universe was about
400,000 years old. The angular power spectrum of the CMB reveals the
initial fluctuation spectrum modulated by the effects of acoustic
waves in the plasma \cite{Hu02}, and it gives quantitative information
about the physical conditions in the plasma. 

The polarization of the CMB provides an independent way to test the
standard
model\cite{Rees68,Kaiser83,Bond84,Bond87,1997NewA....2..323H}.
Anisotropic Thomson scattering of photons at the time of last
scattering gives rise to weak linear polarization of the CMB.
Measurement of the CMB polarization not only provides an additional
way to measure the parameters of the model, it can also verify the
correctness of many of the basic assumptions on which the model is
founded. It is for this reason that many experiments are being
designed to measure the CMB polarization power spectra, despite the
difficulty of the observations.

Angular power spectra give the variance $C_l$ (usually expressed in
terms of CMB temperature, and with units of $\mu$K$^2$) as a function
of multipole number $l$ \cite{Hu02}. The
intensity of polarized radiation can be expressed by the four Stokes
parameters $I$, $Q$, $U$, and $V$ \cite{Stokes1852,Chandrasekhar50}.
Total intensity is represented by $I$, linear polarization by $Q$ and
$U$, and circular polarization by $V$. Thomson scattering does not
generate circular polarization, so we ignore $V$. From the other
parameters we can generate three power spectra $TT$, $EE$, and $BB$,
where $T$ is the total intensity (Stokes $I$) and $E$ and $B$ are the curl-free and
curl-like components of the linear polarization field (Stokes $Q$ and $U$)
\cite{Seljak97,SeljakZaldarriaga97,Kamionkowski97} and also three
cross-spectra $TE$, $TB$, and $EB$.  Because of the parity properties of
the $T$, $E$, and $B$ signals, the only non-zero spectra should be
$TT$, $EE$, $BB$, and $TE$.  In the standard model $E$-modes are
generated from the primary scalar density fluctuations, whereas
$B$-modes are generated only by gravitational-wave tensor fluctuations
and secondary processes;  the predicted $BB$ power spectrum is
undetectable with current sensitivity. A high level of $BB$ would
require modifications to the standard model, or it could indicate that
the observations are contaminated by radiation from foreground
sources, because these are epected to produce both $E$ and $B$ modes in
equal measure.  The polarization spectrum of the CMB is more
difficult to study than the total intensity spectrum because the
fractional polarization of the CMB radiation is no more than $10\%$.
After a number of experiments that placed upper limits on CMB
polarization, $EE$ power has been detected by the DASI [Degree Angular Scale interferometer] experiment
(with $6.3\,\sigma$ significance) \cite{Leitch02,Kovac02,Leitch04} and the
CAPMAP [Cosmic Anisotropy Polarization Mapper] experiment ($2.3\,\sigma$) \cite{Barkats04}. $TE$
cross-spectral power has been detected by DASI and by WMAP
\cite{Bennett03}.

The $TT$ spectrum arises from density and temperature fluctuations
in the plasma, but polarized radiation, which is caused by the
local quadrupole at the time of last scattering, is sensitive to
the velocity of the plasma. Because velocity and density are out of
phase in an acoustic wave, the maxima in the $EE$ spectrum are out of phase
with those in the $TT$ spectra. This phase shift between the spectra is a
key feature of the standard model. It has been seen at large angular
scales in the $TE$ spectrum by WMAP \cite{Kogut03}, but it has not
yet been verified directly through $EE$ or at the small angular scales
corresponding to clusters of galaxies.

We report here observations made with the Cosmic Background Imager (CBI)
that have sufficient sensitivity and resolution to detect and measure
the second, third, and fourth peaks in the $EE$ spectrum \cite{note_peaks},
determine the $TT$-to-$EE$ phase shift, and thus further test the
standard model.

\section*{The Cosmic Background Imager}
        
The CBI (Fig.~\ref{fig:cbi}) has been making observations of the CMB
from a site at 5000~m elevation on the Chajnantor plateau in the
Chilean Andes since late 1999.  It is a 13-element radio
interferometer receiving radiation in 10 1-GHz frequency channels
covering 26 to 36 GHz \cite{Padin02,cbiwebsite}.  The individual
antennas are 0.9~m in diameter, and the possible baselines range in
length from 1.0 to 5.5~m. An interferometer baseline of length $d$ is
sensitive to multipoles $l$ around $2\pi d/\lambda$ where $\lambda$ is
the observing wavelength. The CBI can thus measure the spectrum from
$l\sim300$ to $l \sim 3500$.  The antennas are mounted on a platform
with azimuth and elevation axes that allow all the antennas to track a
point on the sky. The platform can also be rotated about the line of
sight; this allows full sampling of all possible baseline orientations
and facilitates calibration of the instrumental polarization effects.

Each antenna is sensitive to a single sense of circular polarization,
right ($R$) or left ($L$).  Co-polar baselines, $RR$ and $LL$, are
sensitive to Stokes $I\pm V \approx I$ (assuming circular polarization
is negligible), whereas cross-polar baselines, $RL$ and $LR$, are
sensitive to linear polarization, Stokes $Q\pm iU$ \cite{Cotton99}. It
was thus straightforward to adapt the CBI to measure linear
polarization by changing the sense of some of the antennae to maximize
the number of cross-polar baselines [Supporting Online Material (SOM) Text].


The observations reported here were carried out between 22 September
2002 and 7 May 2004 using 7 antennae with polarizers set to left
circular polarization ($L$), and 6 antennae with polarizers set to
right circular polarization ($R$). To avoid contamination by the sun
and moon, observations were made only at night and at angles of
greater than $60^\circ$ from the moon (supporting online text).

\begin{figure}
\includegraphics[width=\textwidth]{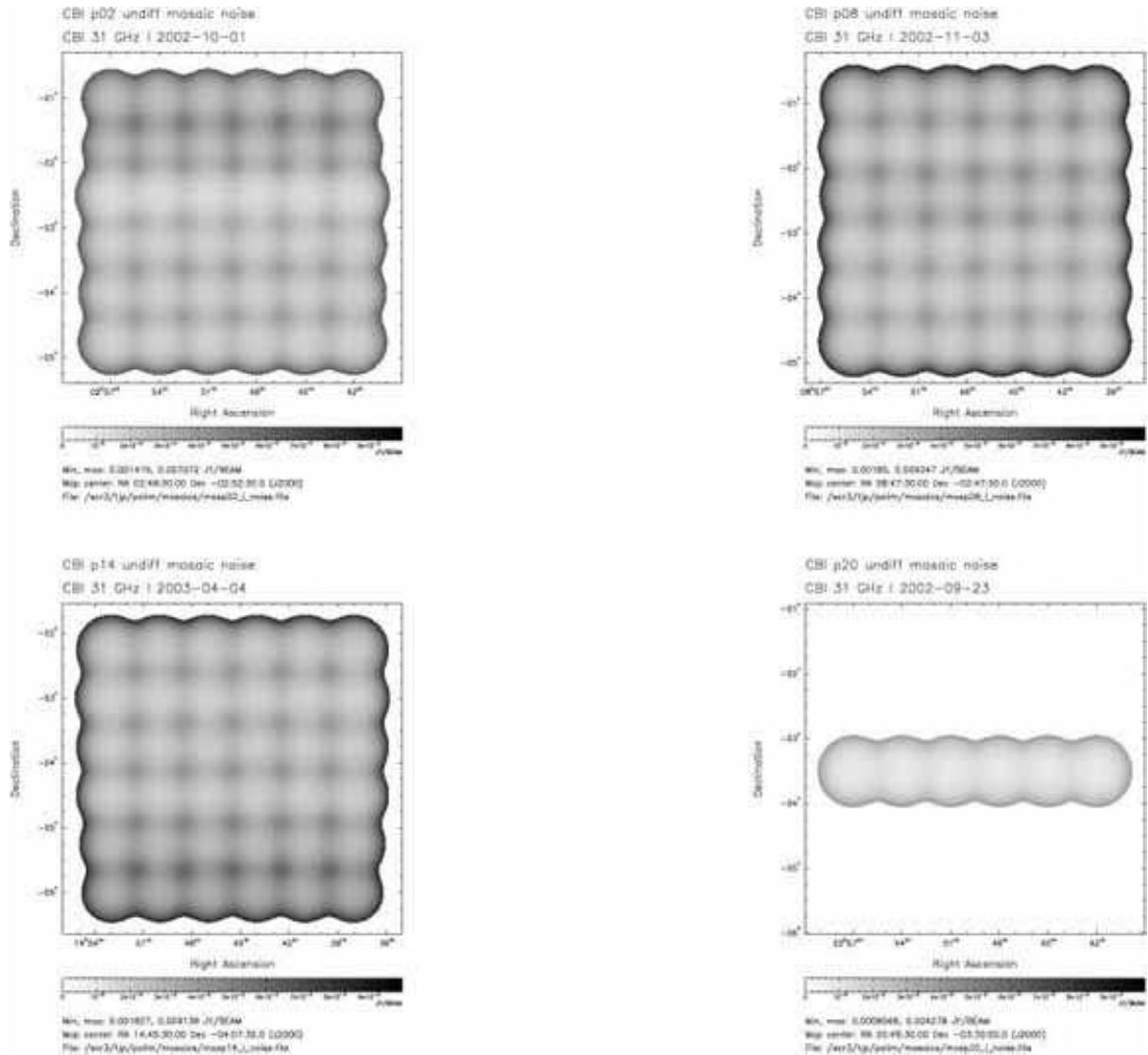}
\caption{The sky coverage of the four fields imaged by the CBI in polarization.
  The grey-scale shows the noise-level achieved in total intensity,
  $I$, in the observations reported here. Three of the fields have
  been mapped with 36 separate pointings, whereas the fourth, $20{^{\rm
  h}}$, has been mapped more deeply but in only six pointings. The
  modulation of the sensitivity by the CBI primary beam is apparent.
  The approximate centers of the four fields are: 
  $02{^{\rm h}}49{^{\rm m}}30{^{\rm s}}$, $-02^{\circ}52'30''$;
  $08{^{\rm h}}47{^{\rm m}}30{^{\rm s}}$, $-02^{\circ}47'30''$;
  $14{^{\rm h}}45{^{\rm m}}30{^{\rm s}}$, $-04^{\circ}07'30''$;
  $20{^{\rm h}}49{^{\rm m}}30{^{\rm s}}$, $-03^{\circ}30'00''$
  (J2000 right ascension and declination).}
\label{fig:fields}
\end{figure}

The size of the CBI antennae sets the $l$ resolution of observations
made in a single pointing to $\Delta l \approx 300$. To obtain the
higher resolution in $l$ necessary to resolve the expected structure
in the $EE$ spectrum, we had to image a larger area by making
a mosaic of overlapping pointings. From 2002 to 2004,
we observed a grid of pointings in four regions near the celestial
equator that were separated by about 6 hours in right ascension and identified
as the $02^{\rm h}$, $08^{\rm h}$, $14^{\rm h}$, and $20^{\rm h}$
fields (Fig.~\ref{fig:fields}).  These fields were centered on those
we observed in 2000 and 2001 in order to measure the $TT$ power spectrum
\cite{Padin01,Mason03,Pearson03,Myers03,Sievers03,Bond02,Readhead04}.
The separation of the pointings was 45~arcmin, twice that of the
earlier observations, leading to modulation of the sensitivity across
the field.  For three of the fields we used 36 different pointing
positions giving fields $\approx5^\circ$ square, but for the 20$^{\rm
  h}$ field we divided the available integration time between six
pointings in a row; these deeper observations should be more sensitive
to any potential systematics.

\begin{figure}
\includegraphics[width=\textwidth]{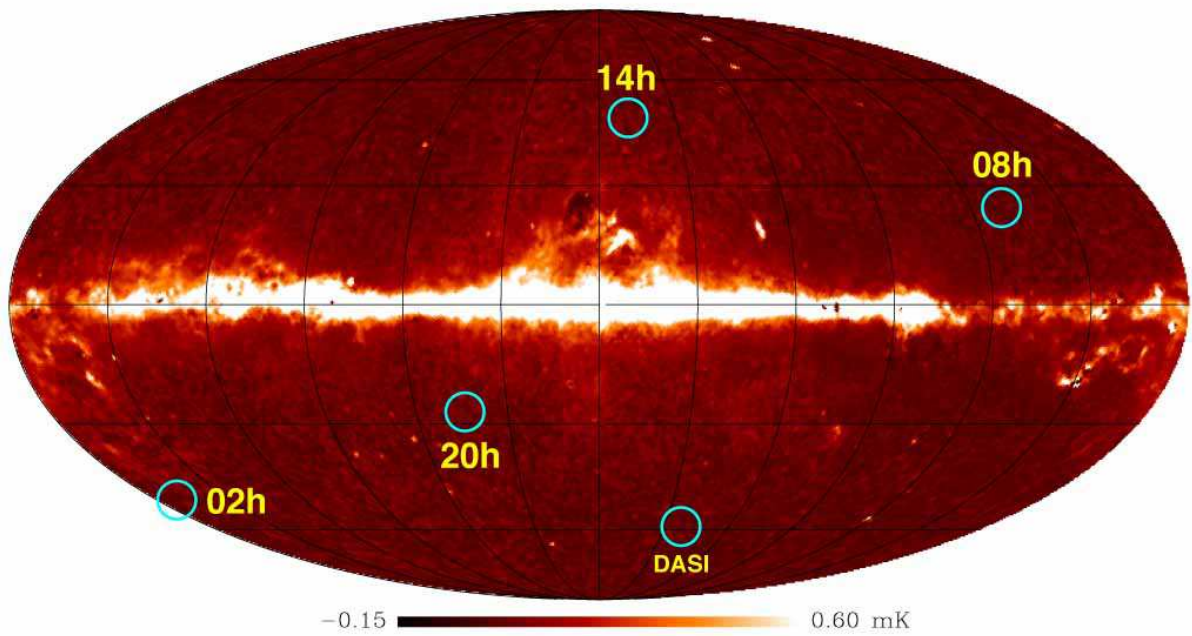}
\caption{Location of CBI and DASI fields in relation to the Galaxy. The
  sky image is the Ka-band synchrotron map derived from {\it WMAP} first 
  year data \cite{Bennett03fg}. Galactic longitude increases to the left,
  with zero in the center of the image.}
\label{fig:galaxy}
\end{figure}

The largest source of diffuse foreground contamination over the 26- to
36-GHz band is synchrotron radiation from the Galaxy.  We
chose the four CBI fields, which were constrained to be separated by
about 6 hours in right ascension, so as to minimize this contamination
(Fig.~\ref{fig:galaxy}).  The CBI $02^{\rm h}$, $08^{\rm h}$, and
$20^{\rm h}$ fields, like the DASI polarization field, are in regions
of low synchrotron emission (Fig.~\ref{fig:galaxy}), but the $14^{\rm
  h}$ field is near the North Polar Spur, and the WMAP observations
suggest that this has a higher level of synchrotron foreground.  The
wide separation of the CBI fields provides some control on foreground
contamination, because foreground emission is unlikely to be
correlated over such large distances and if it were a problem we would
expect to see  differences between the spectra of the
different fields.

The largest systematic instrumental effect we have to eliminate is
ground spillover.  Although the ground radiation is unpolarized it
enters the CBI feeds after reflection off the inner surface of the
shield cans that reduce cross-talk between receivers \cite{Padin02}
and therefore gives rise to a highly polarized contaminating signal.
This signal is particularly strong on the shortest baselines at the
lowest frequencies. To eliminate the ground radiation, we observed
sets of six fields separated by 3~min in right ascension at the same
azimuths and elevations, spending 3~min on each field, so that if the
ground emission is constant it should make equal contributions to all
six fields. When estimating the power spectrum we made use only of the
differences between the fields, ignoring the contribution that is
common to all six. This strategy requires that the ground be stable
over the total scan duration of 18 minutes, but the penalty is only
$\sqrt{6/5}$ in flux density sensitivity, equivalent to a factor of
1.2 in observing time.

The data were edited to remove data corrupted by instrumental or
atmospheric problems. Amplitude calibration was based on Jupiter
\cite{Readhead04} and polarization position-angle calibration was
based on Tau A, for which we measured a polarization position angle
($E$ vector) of $-27.6^\circ$ by comparison of CBI and Very Large
Array \cite{VLA} observations of 3C\,273 and 3C\,279. Instrumental polarization
leakage was measured on Tau~A and found to be negligible. The noise
was estimated from the scatter of the measurements in each 18-minute
scan (SOM Text).

\begin{figure}
\includegraphics[width=\textwidth]{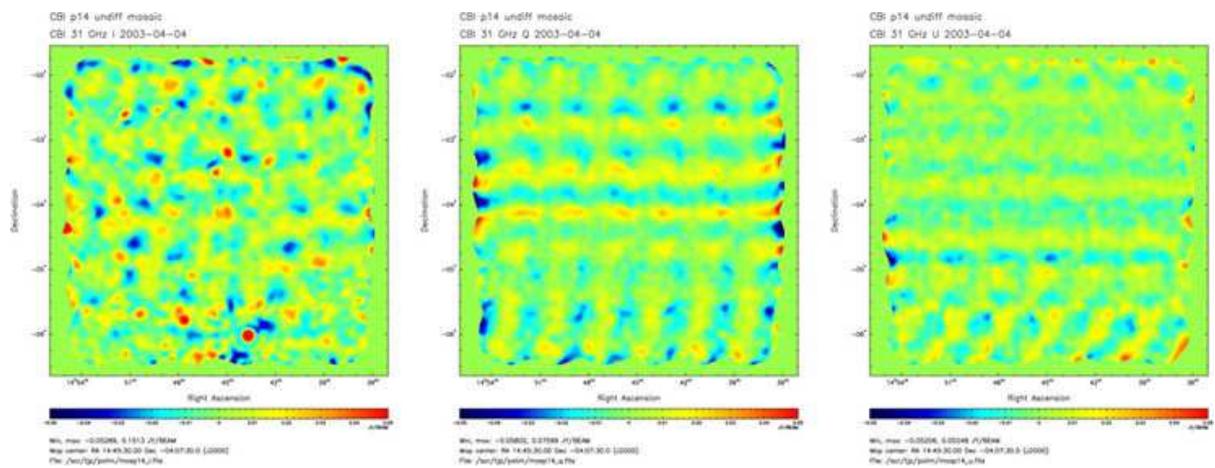}
\caption{Images of the 14$^{\rm h}$ field mapped by the
  CBI in Stokes parameters $I$, $Q$, and $U$ (Stokes $V$, circular
  polarization, is not measured and is assumed to be zero). Color is used
  to represent intensity, with the same scale in each Stokes parameter.
  In these images the contaminating effects of ground radiation and
  foreground emission have not been removed.  The total intensity,
  $I$, image ({\it left}) is dominated by CMB emission (modulated by
  the instrumental point-spread function); some foreground point
  sources are visible ({\it red spots}). The linear polarization, $Q$
  and $U$, images ({\it center and right}) are dominated by
  instrumental noise and ground pickup. Ground pickup, which with our
  observing strategy should be the same in each pointing at the same
  declination, gives rise to a pattern that repeats at intervals of
  3~min in right ascension.}
\label{fig:images}
\end{figure}

\begin{figure}
\includegraphics[width=\textwidth]{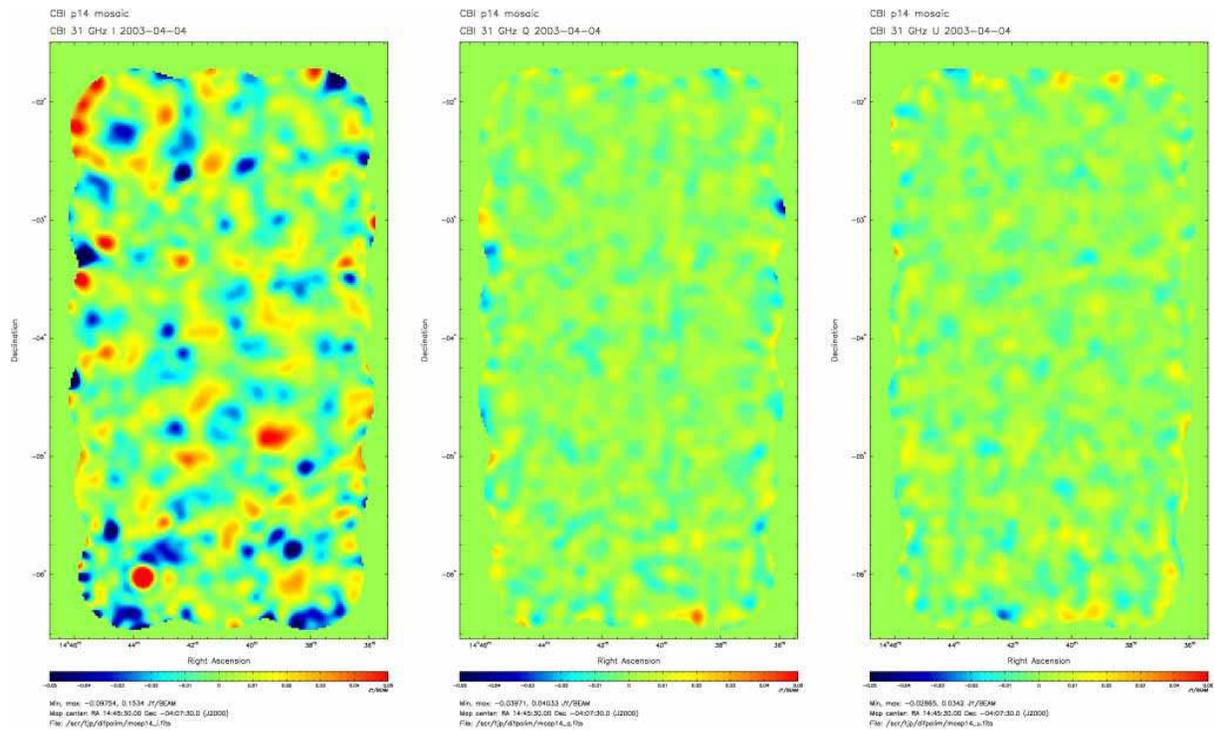}
\caption{The effect of lead {\it minus} trail differencing. 
  Here the data presented in Fig.~\ref{fig:images} have been
  differenced: each visibility measurement has had the corresponding
  measurement on a field 9~min later in right ascension subtracted. Because
  the ground pickup is very similar for both measurements, ground
  emission cancels out in the difference. In the resulting images
  foreground point sources may appear positive or negative in $I$ depending
  on their right ascension. The $Q$ and $U$ images show that ground
  pickup has been removed with high accuracy.}
\label{fig:difimages}
\end{figure}

After the editing and calibration of the data we made images of
the four fields and of all the calibration sources in order to check
for possible anomalies \cite{Pearson03}.  As an example, the $I$ image
of the 14$^{\rm h}$ field (Fig.~\ref{fig:images}), made without any
subtraction of ground spillover or foreground sources, shows
significant power above the level of the noise.  This is due to both
CMB emission and ground spillover.  In the $Q$ and $U$ images
(Fig.~\ref{fig:images}) the signal from the CMB is too weak to
identify, and these images are dominated by the regular pattern due to
ground spillover.  The level of the ground spillover in $Q$ and $U$
indicates that there is some ground contamination in the $I$
images as well, although it is somewhat weaker than the CMB signal.
When we estimate power spectra, the ground spillover is removed from
the data by projecting out \cite{bjk98,Halverson02} the common mode in
the six matched pointings, so the visibility data set from which the images
of Fig.~\ref{fig:images} were made is the data which we use in the 
CMB spectrum determination.  But in order to check our procedures we
have also made images from the differences of visibilities
measured in pointings separated by 9 min in right ascension; these
images should be free of ground contamination
(Fig.~\ref{fig:difimages}). The total intensity $I$ image shows power
well in excess of the noise level, whereas the $Q$ and $U$ images show
only noise, the sensitivities per resolution element being too low to
reveal the polarization of the CMB.  These images also show that
leakage of total intensity into the polarization data is small
compared to the noise.

\section*{Power Spectrum Estimation}

To estimate power spectra from the interferometer visibility
measurements we used maximum-likelihood procedures similar to those
adopted for earlier experiments \cite{White99,Kovac02,Park03}. To
process the CBI data, we have extended the gridding-based procedure
used in our earlier work \cite{Myers03} to deal with mosaicked
polarization observations.  A given correlator output sample, or
visibility, can be one of the four polarization products $RR$, $RL$,
$LR$, or $LL$.  These can be related to the fundamental CMB
polarization modes $T$ (temperature), $E$, and $B$ (polarization)
\cite{Kovac02}.  The covariances between the measurements depend on
the six CMB covariances $TT$, $EE$, $BB$, $TE$, $TB$, and $EB$.
Because the CBI measures circular polarization products, which are
orientation independent (depending only on the
handedness of the wave polarization), the CBI (or any interferometer
using circularly polarized receptors) is sensitive to the $E$ and $B$
modes directly.  This simplifies the power spectrum analysis
(SOM text).


The principal foreground contamination in total intensity for the CBI
is that due to extragalactic radio sources \cite{Mason03}.  For the total
intensity spectrum, in which discrete sources have a substantial impact,
our approach is similar to that used in earlier CBI analyses
\cite{Mason03,Pearson03,Readhead04,Myers03},
with minor modifications.  Some 3727 NRAO Verly Large Array (VLA) Sky Survey (NVSS) \cite{NVSS} sources with
$S_{1.4\,{\rm GHz}} \ge 3.4 \, {\rm mJy}$ were projected out of the
data.
In previous
work we used separate covariance matrices with different projection
factors for sources that were detected at 32~GHz on the OVRO
40-meter telescope and for those that were not. Because in the end there
was no gain from this approach, in the present analysis we
combined all sources into a single covariance matrix with a single
projection factor.  For this analysis we assume a uniform variance of
1~Jy$^2$ for each source, rather than adjusting the variance for each source.
We find that this yields matrices that
are numerically more stable under the action of our procedure of
completely projecting the source modes out of the data.  After a
number of tests, we adopted a value of $q_{\rm src}=100$ for the
pre-factor (equivalent to setting the variance on each source flux
density to 100~Jy$^2$).

However, because non-thermal extragalactic radio sources are weakly
polarized ($P\le 10\%$) and furthermore only a small fraction of them
have $P>2\%$, only a few of the sources that we projected out in the
total intensity spectrum can affect the polarization spectrum.  When
estimating $EE$ and $BB$, we therefore projected out only a subset
of the NVSS sources with $S_{1.4}>3.4$ mJy.  In total  556 of
these potentially troublesome point sources need to be considered.
These include (i) NVSS sources with $>3\sigma$
detections of polarized flux density at $1.4 \, {\rm GHz}$ and (ii)
sources detected by the 30-GHz OVRO survey of the 2000-to-2001 CBI total
intensity fields \cite{Readhead04}.
The projection of 556 sources out of the CBI data has only a small
effect on the $EE$ power spectrum: in all bins the effect is $\ll
1\sigma$. In the first two bins, where the polarization detection is
strongest, the effect is less than $3 \, {\rm \mu K^2}$ for each bin.
Both with and without projection, the $BB$ power spectrum is consistent
with zero, and the $EE$ spectrum changes very little. Therefore, the
sources that we have identified as potentially troublesome (with the
criteria described above) have a negligible effect.  If we have failed
to identify some sources (highly polarized sources just below the NVSS
detection limit, for instance), they should add a characteristic
$l(l+1)C_{l} \propto l^2$ contribution to both $EE$ and $BB$ power
spectra, and show up more strongly in the lower frequency channels. No
such signature is evident in the CBI data.

We have also studied the effects of point sources using Monte Carlo
simulations.  To do so we used the NVSS source statistics to
characterize the fractional linear polarization of sources, finding a
mean $1.4 \, {\rm GHz}$ polarization of $2.7 \%$. Most sources had
polarizations less than this; $4\%$ had polarizations greater than
$10\%$, and $1\%$ had polarizations greater than $15\%$. Because the
fractional polarization of CMB anisotropies is  $\sim
10\%$, the
discrete source foreground will be relatively weaker in
polarization than in total intensity.  In the $EE$ power
spectrum analysis of the simulated data, we find that the first two
bins change by less than $4 \, {\rm \mu K^2}$ when the source
projection is turned on, similar to what is seen in the real
CBI data.


The common-mode signal from the ground was removed by constructing a scan
covariance matrix assuming a unity correlation between identical visibilities
coming from the same scan (e.g., for the six visibilities taken in the
consecutive 3 minute integrations that constitute a scan), then passed through
the gridding operation.  The modes defined by this scan covariance matrix were
projected out of the data \cite{bjk98} by applying a large prefactor to this
matrix in the likelihood maximization procedure (essentially setting the
variance of these modes to be infinite), in the same manner as for the
point sources. 
We used simulated data to determine the best value of the prefactor, and found
that too small a value did not completely eliminate the ground spillover,
whereas too large a value caused numerical problems.  Because there are a large
number of sources in the list used for the $TT$ projection, there is an
interaction between the source and scan projection matrices $q_{\rm src}\,{\rm
C}^{\rm src}$ and $q_{\rm scan}\,{\rm C}^{\rm scan}$ when the pre-factors
$q_{\rm src}$ and $q_{\rm scan}$ become large.  We explored a range of values
for these, and found that for the value of $q_{\rm src}=100$ a
value $q_{\rm scan}=100$ was in the center of the range for which the $TT$
band powers were stable (there was no significant change in $TT$ amplitude from
$q_{\rm scan}=10$ to $q_{\rm scan}=100$).  Similar tests on the real data
also showed that the ground signal was eliminated whereas the band powers
remained stable.

\section*{Polarization Power Spectra}
        
\begin{figure}
\includegraphics[width=\textwidth]{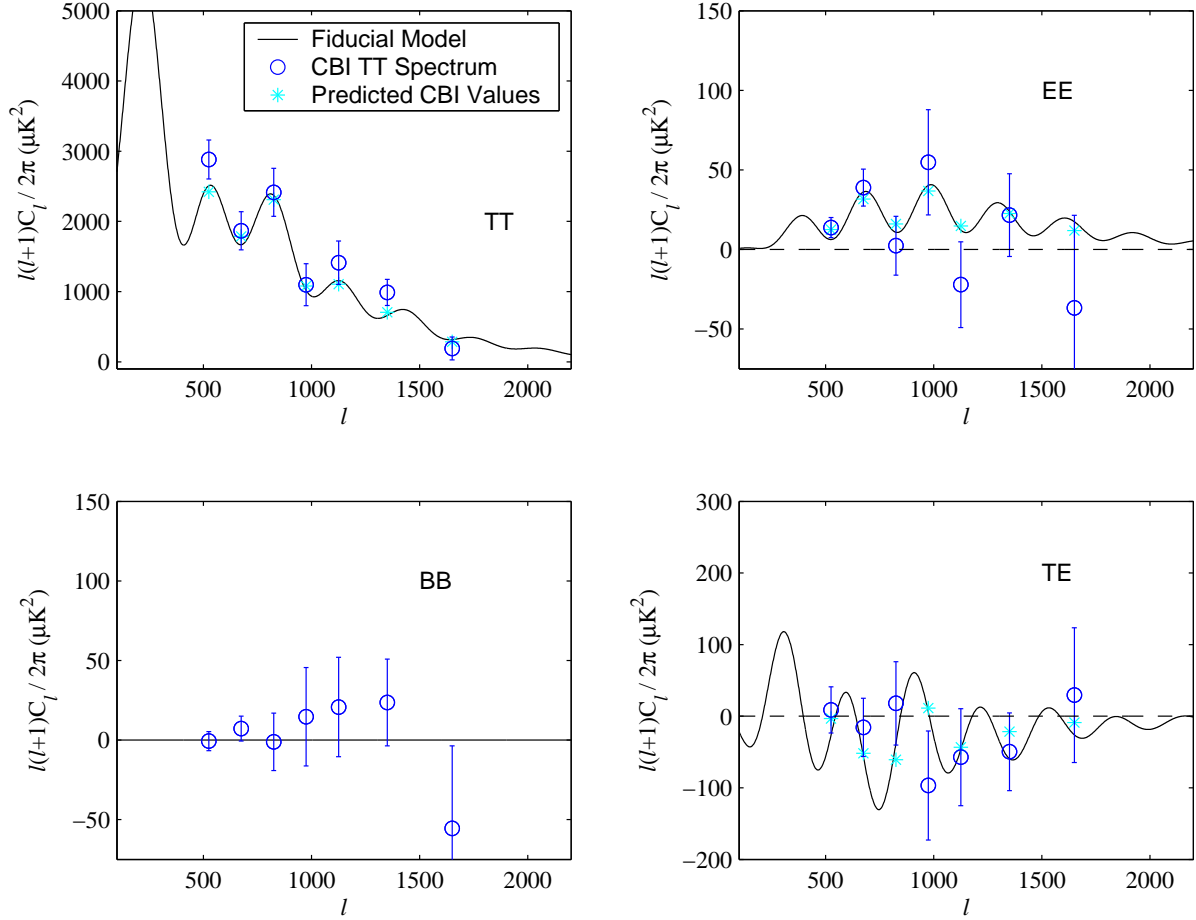}
\caption{Power spectra of CMB polarization from the CBI measurements.
  The four panels show total intensity power spectrum $TT$, grad
  polarization mode power spectrum $EE$, curl
  polarization mode power spectrum $BB$, and cross-spectrum $TE$.  Numerical
  values are given in Table~\ref{tbl:powers}. The {\it black curve} 
  is the theoretical
  $\Lambda$CDM model using a power law for the primordial spectral
  index which best fits the {\it WMAP}, CBI, and ACBAR CMB data \cite{wmapmodel}. The predictions of this model
  for the CBI bands using the CBI window functions are indicated by
  the {\it stars}.}
\label{fig:spectra}
\end{figure}

\begin{figure}
\includegraphics[width=\textwidth]{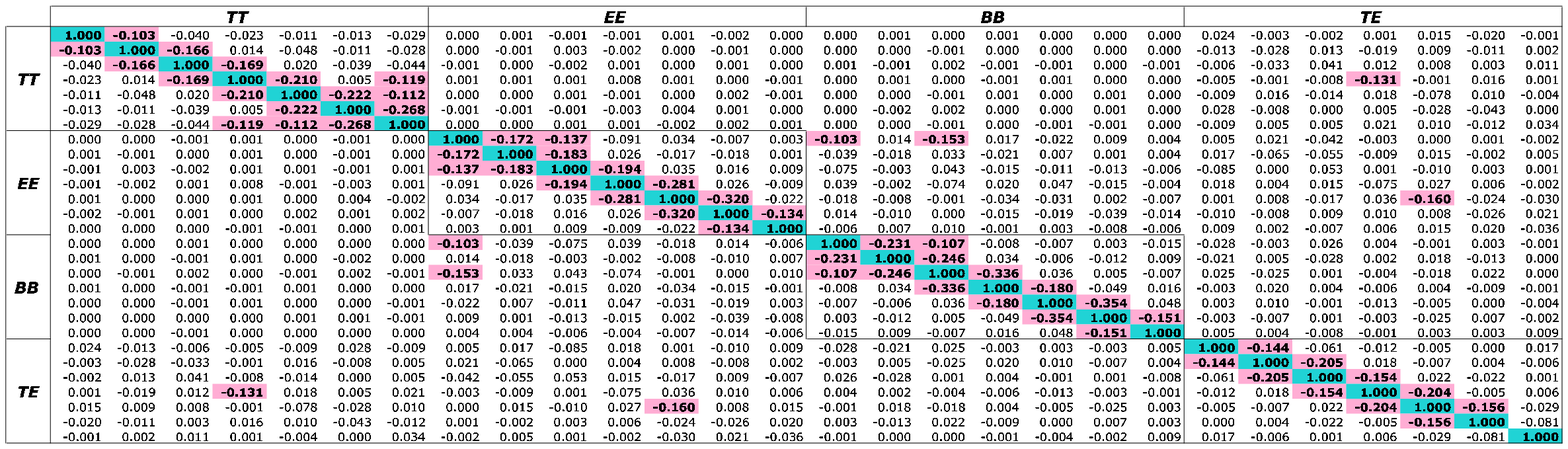}\\
\caption{Normalized band-to-band correlations for the 7 bands shown in Table~\ref{tbl:powers}.}
\label{fig:correlations}
\end{figure}

\begin{figure}
\includegraphics[width=0.45\textwidth]{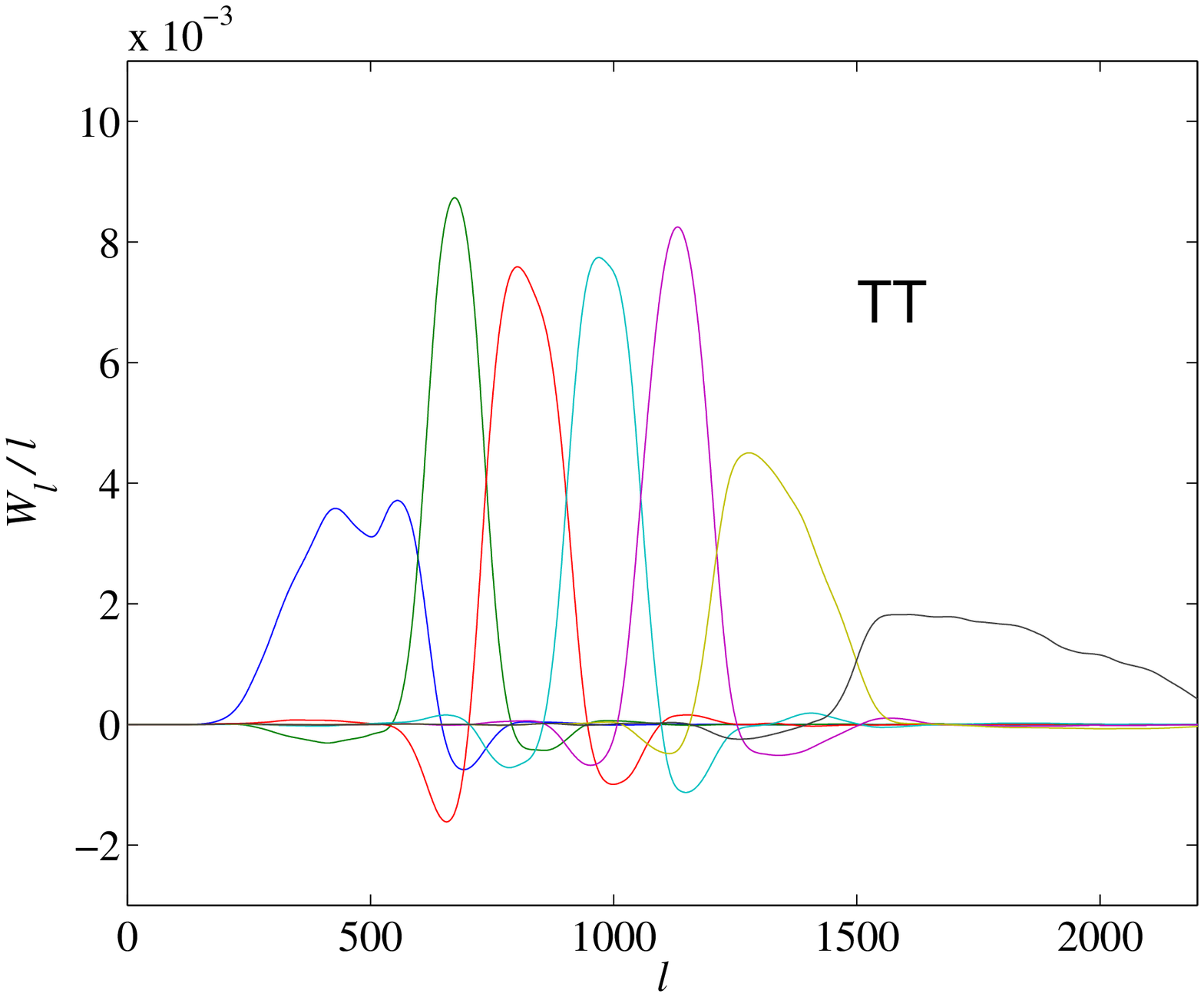}
\includegraphics[width=0.45\textwidth]{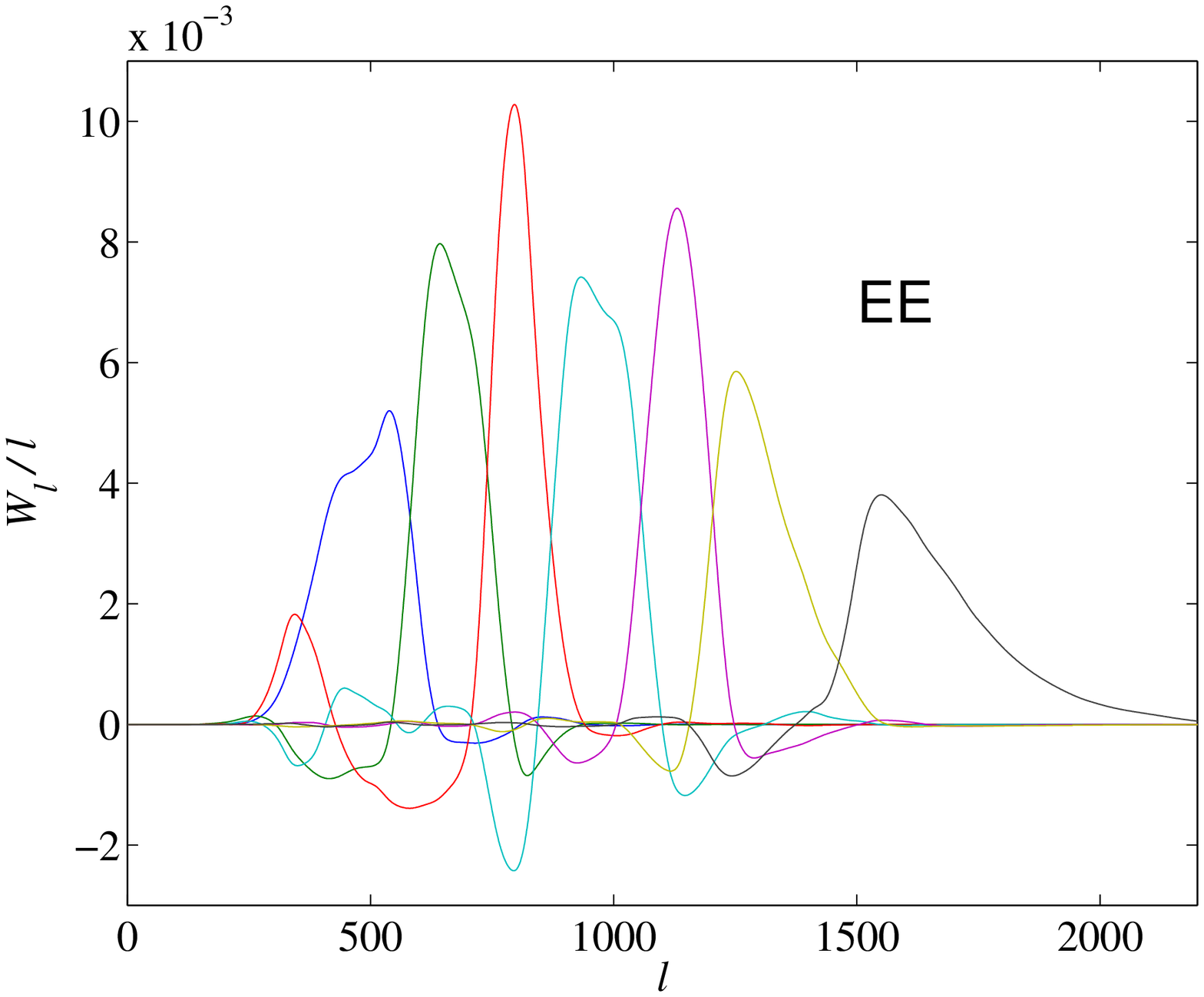}\\
\includegraphics[width=0.45\textwidth]{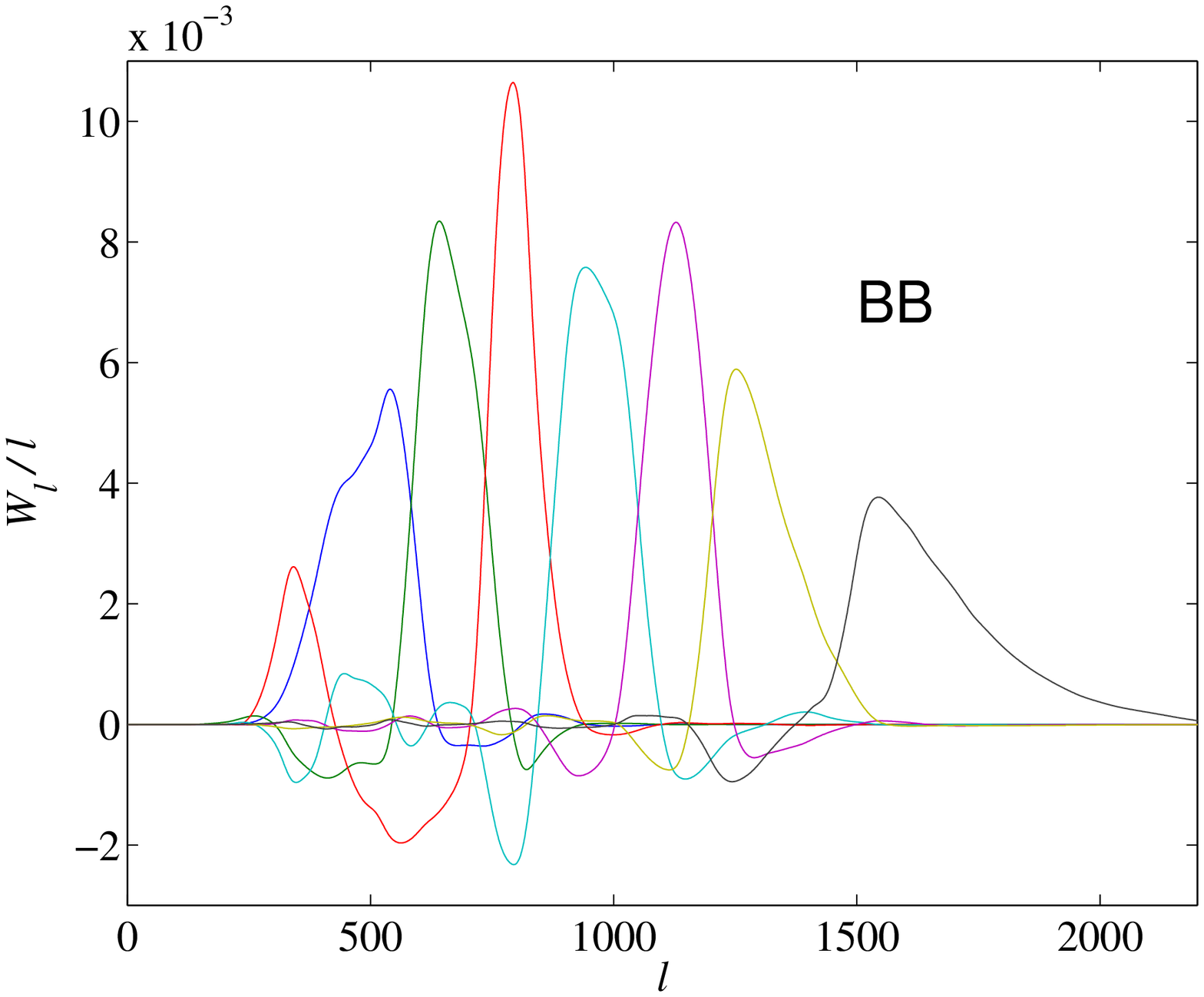}
\includegraphics[width=0.45\textwidth]{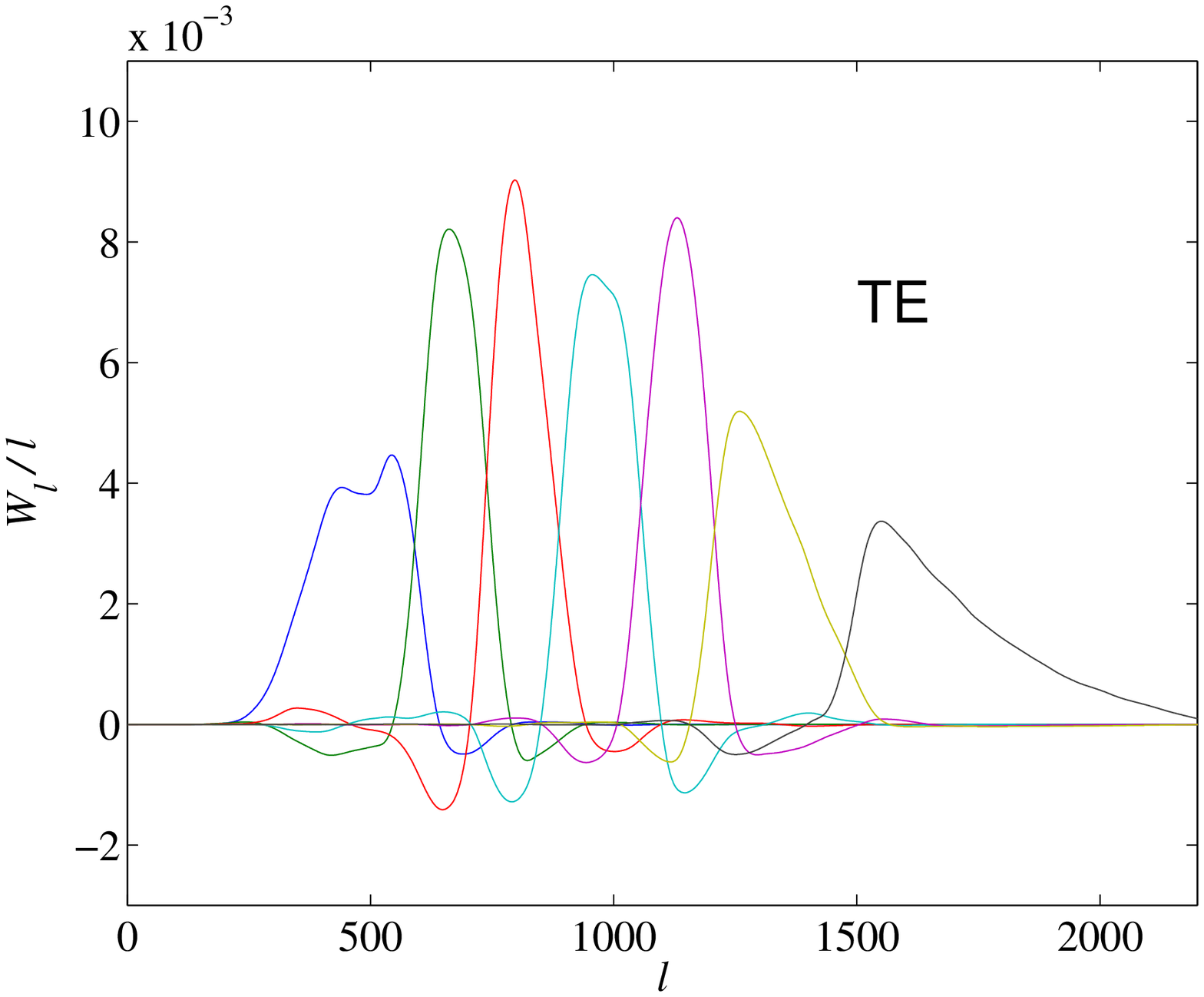}\\
\caption{Band-power window functions for the CBI polarization observations with
  the bands shown in Table~\ref{tbl:powers}.  The expected value of
  band power for a given model spectrum, $C_l$, is $\sum_l
  \left[W(l)/l\right]l(l+1)C_l/2\pi$.  The total area under each window
  function is equal to unity.  The band powers for each spectrum also
  contain contributions from the the others ($EE$ band powers are
  affected by changes to the model $TT$ and $TE$ as well as $EE$, for
  example), but for
  the CBI these cross-polarized window functions are small, with peak
  amplitudes of no more than a few percent of the same-polarized
  window functions.}
\label{fig:windows}
\end{figure}

The CBI measurements for all 10 frequency channels and all four fields
observed from 2002 to 2004 have been combined in the maximum likelihood
procedure to estimate the $TT$, $EE$, $TE$ and $BB$ power spectra
(Table~\ref{tbl:powers} and Fig.~\ref{fig:spectra}).  The scan means (for
ground contamination) and point sources have been projected out as
described in the previous section.  We divided the $l$ range into seven
bands, with most of the bins having width $\Delta l = 150$. Adjacent
bands are anticorrelated at the 10 to 20\% level
(fig.~\ref{fig:correlations}). Finer binning is possible, but this
gives larger band-to-band correlations and is less satisfactory for
presentation.  For the quantitative analysis below, we have used bins
of width $\Delta l \approx 75$ and taken into account the band-to-band
correlations. The results from both binnings are consistent. We have also calculated window functions that can be used to calculate the expected band powers in our bands from a theoretical model spectrum (fig.~\ref{fig:windows}).

We compare our results with a fiducial model spectrum
(Fig.~\ref{fig:spectra}), for which we have chosen the theoretical
model \cite{wmapmodel} with a power law for the primordial spectral
index which best fits the first-year {\it WMAP}, 2000 CBI, and ACBAR
[Arcminute Cosmology Bolometer Array Receiver] CMB total-intensity
data [the ``WMAPext'' data set \cite{Spergel03}].  Our results are
consistent with the predictions of this model.  We have checked this
by calculating $\chi^2$ for a comparison of our measured band powers
and the band powers predicted by the model, with the CBI window
functions and the full band-to-band covariance (estimated from the
Fisher matrix). The values of $\chi^2$ [for 7 degrees of freedom (df)],
with the probabilities of obtaining larger values under the null
hypothesis in parentheses, are: for $TT$, 7.98 (probability = 0.33); for
$EE$, 3.77 (probability = 0.80); for $BB$, 4.33 (probability = 0.74); and for $TE$, 5.80
(probability = 0.56).

The $TT$ spectrum shows the same features that we saw in the CBI
2000-to-2001 observations \cite{Readhead04}, the most prominent being
the drop in power between the third and fourth acoustic peaks.  The
$TT$ spectrum from 2002--2004 is slightly higher than the fiducial
model, but the difference is not significant.  Both the $EE$ and $TE$
spectra are consistent with the predictions of the fiducial model. The
$EE$ spectrum shows detection of power at $l<800$, whereas the $TE$
spectrum is not sufficiently sensitive to show a positive detection.
No power is detected in the $BB$ spectrum, as expected on the standard
model.  The 95\% confidence upper limit on $BB$ power (assuming flat
band power in a single $l$ bin) is 7.1~$\mu$K$^2$ with source
projection or 2.7~$\mu$K$^2$ without source projection.  Because
ground radiation and foreground sources are expected to contribute
equally to $EE$ and $BB$, this low limit on a possible $BB$ component
at multipoles $l<1000$ demonstrates that there is no significant
ground or point source contamination at these multipoles and gives
confidence in the reliability of the $EE$ spectrum.


At present, the addition of polarization data to CMB $TT$ data has
little effect on the values and precision of cosmological parameter
estimates, because of the weakness of the polarized signal relative to
the total intensity signal. Rather, the strength of the
measurement of $EE$ lies in its ability to test a different aspect of
the theory. It is nonetheless interesting to explore the effect of the
new polarization results on cosmological parameter estimation and to
check consistency. As sensitivities improve, future polarization data
will have a bigger impact on parameters \cite{Bond04}.  In this initial
investigation including CMB $EE$ polarization data, we explore
a limited set of cosmological parameters that has been successful in
describing all aspects of CMB data. The model has its basis in the
simplest inflationary paradigm, characterized by the following basic
set of six parameters: $\omega_b\equiv\Omega_bh^2$, the physical
density of baryons; $\omega_c\equiv\Omega_ch^2$, the physical density
of cold dark matter; $\theta\equiv100\ell_s^{-1}$, parameterizing the angular scale
$\ell_s^{-1}$ associated with sound crossing at decoupling, which defines
the overall position of the peak--dip pattern; $n_s$, the spectral
index of the scalar perturbations; $\ln A_s$, the logarithm of the
overall scalar perturbation amplitude; and $\tau_c$, the Thomson
scattering depth to decoupling. We do not consider any gravitational-wave
induced components because they are not expected to be detectable by the CBI.

The strongest prior we impose is that we only consider flat models
($\Omega_{\rm tot}=1$), as expected in most inflation models. We note that
the parameters we derive can change significantly when this prior is
relaxed  \cite{Bond03,Readhead04}. We also impose a weak-$h$
prior comprising limitations on the parameter $h$ ($0.45 < h < 0.90$, where $h \equiv H_0/(100\,\rm km\,s^{-1}\,Mpc^{-1})$ and $H_0$ is the Hubble constant)
and on the age of the universe ($t_0 > 10 \, {\rm Gyr}$). Within the context
of flat models the weak-$h$ prior influences the results very
little. We do note, however that extreme models with high Thomson
depth are excluded by this prior. 

In our analysis we consider three combinations of data: (i) WMAP1 only
$TT$ and $TE$ results from the first year WMAP data \cite{Bennett03},
using the likelihood procedure described in \cite{Verde03}; (ii) CBI
pol plus WMAP1 obtained by adding 14 band powers in each of $TT$, $TE$,
and $EE$ obtained from an analysis of the 2002-to-2004 CBI data with a
bin width $\Delta l \approx 75$; and (iii) CBI pol plus CBIext plus WMAP1,
consisting of all of the CBI polarization data and the addition of
high-$\ell$ bands from our combined mosaic and deep field $TT$ results
\cite{Readhead04}, covering the range $\ell=600$ to $\ell=1960$ [bands
5 to 14 of table~1 of \cite{Readhead04}]. This third combination
extends the data well into the region known as the damping tail, where
power is suppressed by photon diffusion and the finite thickness of
the last scattering surface.

We use a modified version of the Markov chain Monte
Carlo (MCMC) package COSMOMC \cite{Lewis02,COSMOMC} to evaluate
probability distributions of the various parameters with respect to
the CMB data. We have extended our earlier procedures
\cite{Readhead04,Bond03} to include polarization spectra and the cross
correlation between $TT$ and $EE$ spectra \cite{cluster}.
In addition to estimating the six cosmological parameters defined
above, we determined the distributions of six other derived parameters
from the same Markov chains: $\Omega_\Lambda$ the energy density in a
cosmological constant in units of critical density, the total age of
the universe in Gyear; the total energy density of matter, $\Omega_m$;
the present-day RMS mass fluctuation on $8 \, h^{-1} \, {\rm Mpc}$
scales, $\sigma_8$; the redshift of reionization, $z_{re}$ (related to
$\tau_c$ and $\Omega_b$); and lastly the Hubble parameter $H_0$ in
units of km~s$^{-1}$~Mpc$^{-1}$. $\Omega_\Lambda$ is a derived
quantity determined from $\theta$. The amplitude
parameter $\sigma_8$ is related to $\ln A_s$ and has  more relevance for comparison with large scale structure data.  As expected the inclusion of our
polarization results does not have a large impact for this limited
parameter set (Table~\ref{tbl:paramtab}). However when including the
CBIext $TT$ band powers we obtain significant reduction in the
uncertainties, in  agreement with \cite{Readhead04}.

\paragraph*{Significance of detection.} 
Our standard CMB power spectrum analysis \cite{Myers04} involves the use of
a fiducial $C_\ell$ shape against which the band powers are evaluated.
The gridding procedure breaks the power spectrum into top-hat bands in
$\ell$, and thus the (multiplicative) band powers $q_B$ effectively
break the spectrum into piecewise continous bands that 
follow the shape $C_{B\ell}=q_B\,C_\ell$ for $\ell$ within each band
$B$.  The most conservative choice for a shape is the flat
spectrum $C_\ell=2\pi/\ell^2$, but one can use a matched shape
derived from an actual CMB power spectrum and thus optimally check for
deviations from that model.  This also allows the use of wider $\ell$
bands. 
If we use the fiducial model fitted to the WMAPext dataset
(Fig.~\ref{fig:spectra}) as our shape, and only project point
sources from the $TT$ sector, we find for the CBI data in a single
$\ell$ band a maximum likelihood value band power for $EE$ of
$q_B=1.22\pm0.21$ (68\%) with respect to the WMAP-normalized spectrum,
with a value for the log-likelihood with respect to zero of 39.8
(equivalent to an $8.9\,\sigma$ detection, where
$\sigma=\sqrt{2\,\Delta\log {\cal L}}$). This can be compared with the
detection of $6.3\,\sigma$ reported for the DASI 3-year results
\cite{Leitch04}.  Although there is no indication that the polarization of
the foreground point sources is affecting our data, we can also adopt
a conservative approach and project out the subset of the brightest
sources, as described above.  In that case, the best-fit
band power $q_B$ is $1.18\pm0.24$ (68\%) with log-likelihood with respect
to zero of 24.3 (equivalent to $7.0\,\sigma$). This reduction in
significance is due to the increase in uncertainties from the lost
modes in this projection, i.e., the drop in band powers is negligible,
which again suggests that point sources are not a problem in the $EE$
spectrum.
Although we find no evidence for
point sources affecting our $EE$ spectrum, we adopt this more conservative
 value as our estimate of the  significance of our detection.

\begin{figure}
  A\includegraphics[width=0.45\textwidth]{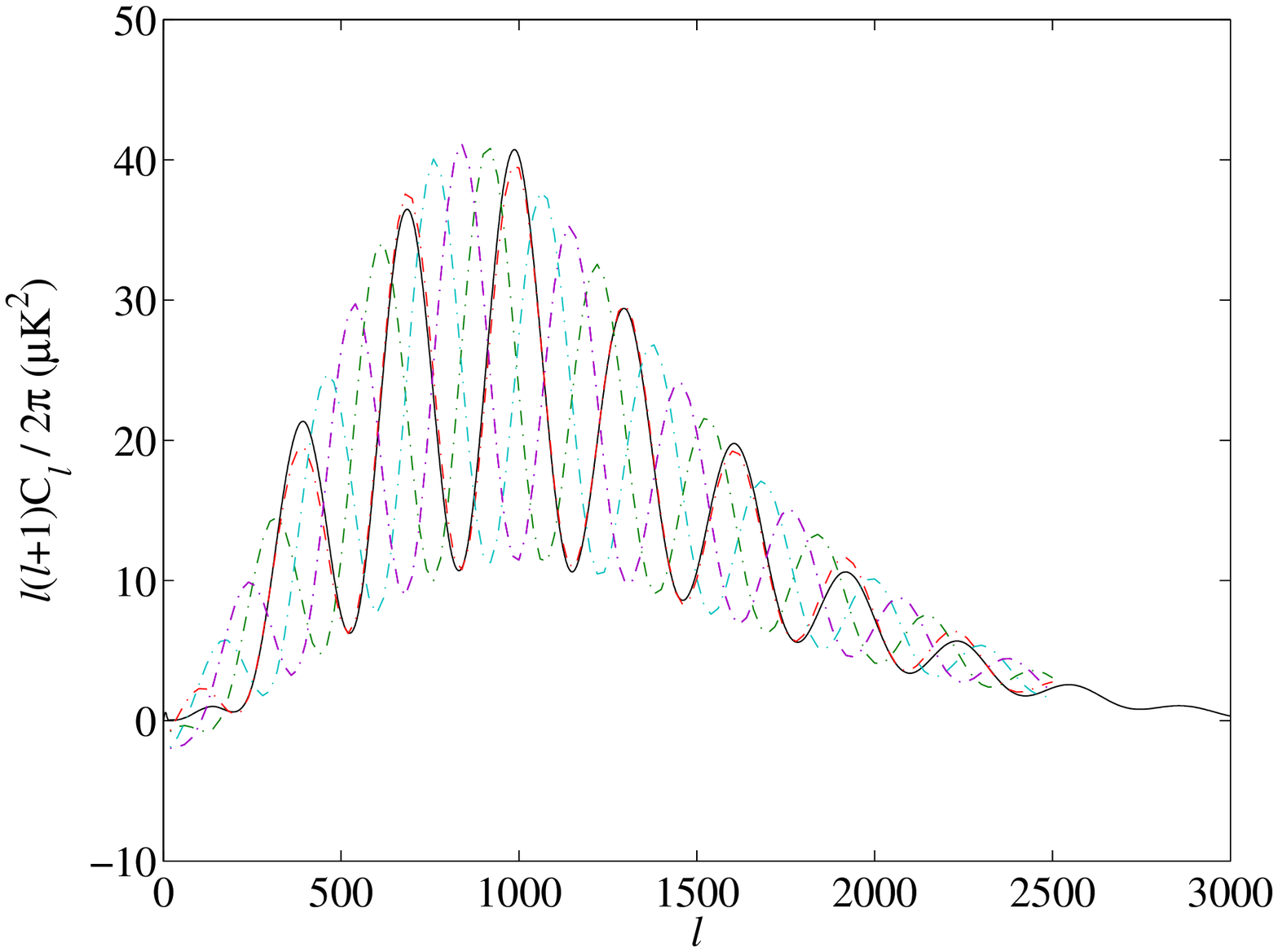}
  B\includegraphics[width=0.45\textwidth]{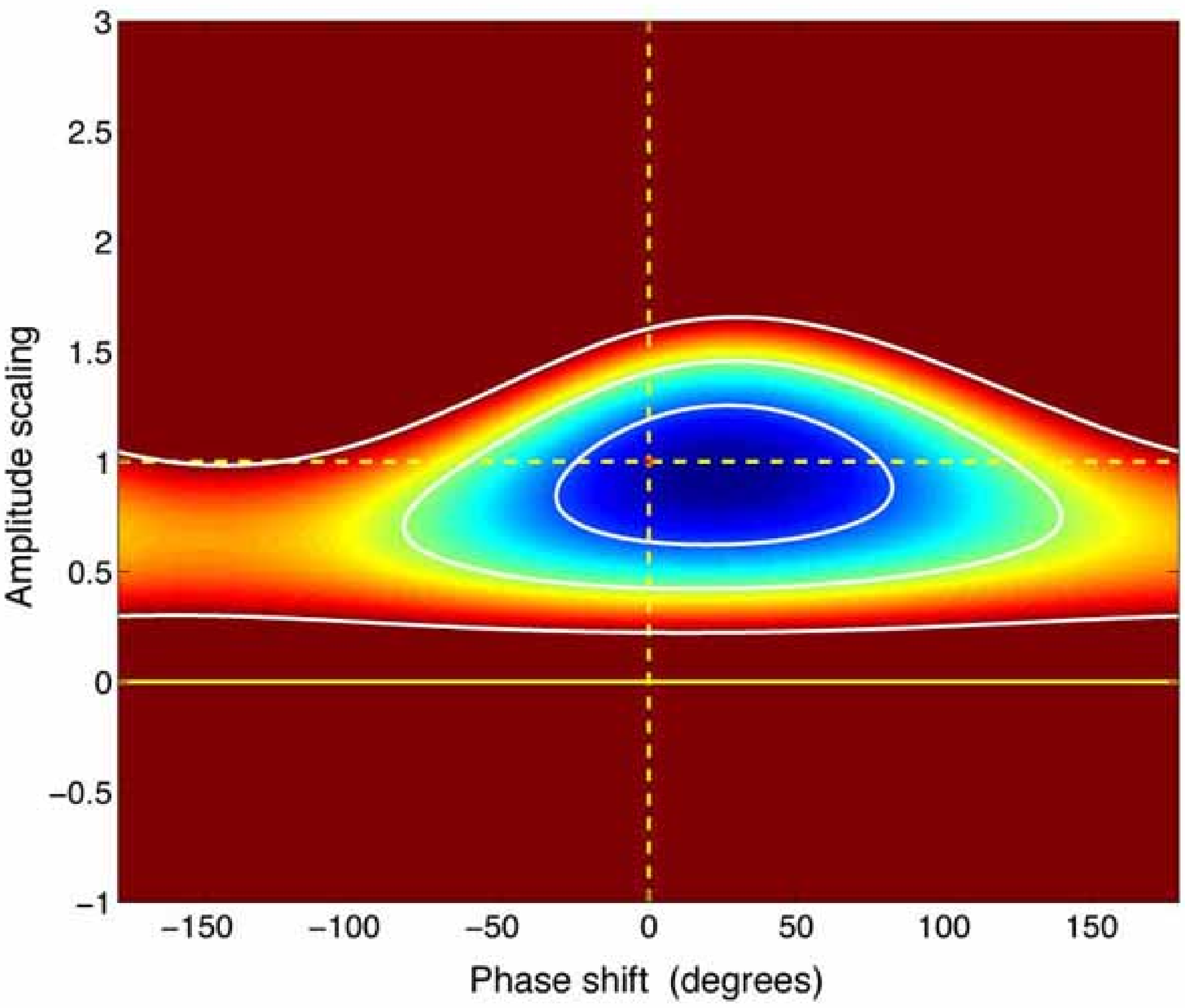}\\
  C\includegraphics[width=0.45\textwidth]{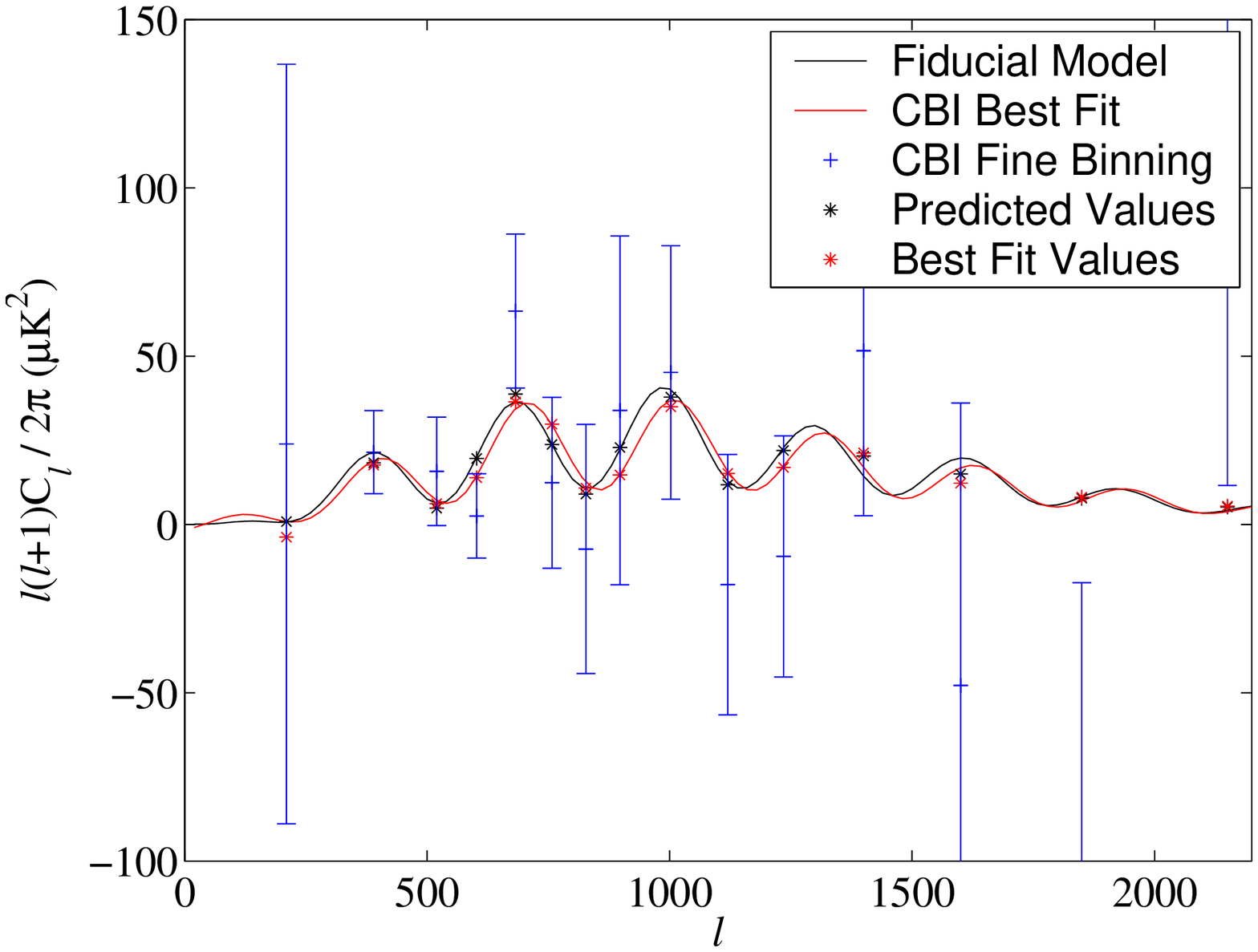}
\caption{Sensitivity of the CBI $EE$ power spectrum to the phase of
  the acoustic oscillations. (A) The $EE$ spectrum predicted by the
  fiducial  model ({\it black line}) with a variety of ``phase
  shifted'' spectra with similar envelopes calculated as described
  in the text. (B) Goodness-of-fit ($\chi^2$) for the model as a
  function of two parameters: the phase shift (horizontal axis) and an
  overall scaling (vertical axis); the point at (0,1) indicates the fiducial
  model corresponding to a standard $\Lambda$CDM model.
  Contours are at 1, 2, and 3 $\sigma$ intervals (i.e., $\Delta \chi^2
  = 2.30, 6.17, $ and $11.8$ for 2 df). 
  (C) Comparison of the fit of the fiducial model ({\it black
  line}) and the minimum-$\chi^2$ phase-shifted model ({\it red
  line}) with the CBI data points ({\it blue)} and the
  band powers predicted by the models ({\it black and red stars}).}
\label{fig:phase}
\end{figure}

\paragraph*{Phase of the acoustic oscillations.}
The measurement of the phase of the polarization $EE$ spectrum can, in
principle, provide one of the fundamental pillars of the standard
model because it tests a unique aspect of the acoustic waves in the
photon-baryon fluid. The peak positions in $TT$ are proportional to $\pi
\ell_s j$, whereas for $EE$ polarization they are proportional to $\pi \ell_s
(j+1/2)$, with some corrections from projection effects.
To test this, we
devised phenomenological models in which the phase-relationship
between $TT$ and $EE$ is changed. For these models,  we first approximated
 the fiducial model $EE$ spectrum as a function:
\begin{equation}
l(l+1)C_l^{EE}=f(l)+g(l) \sin (kl+\phi)
\end{equation}
where $f$ and $g$ are smooth, non-oscillating functions (we used
rational functions with quadratic numerator and denominator) and $k$ is a
constant. We then varied $\phi$ to get a range of phase-shifted
spectra (Fig.~\ref{fig:phase}A). 
To determine
the goodness-of-fit of  the phase-shifted models, we calculated
$\chi^2$ as a function of the phase $\phi$ and a scaling amplitude
$A$, taking into account bin-to-bin correlations using the inverse
Fisher matrix (Fig.~\ref{fig:phase}B). For
this exercise we used the $\Delta l \approx 75$ binning of the CBI
power spectrum.  The best-fit phase is $24^{\circ}\pm33^\circ$ with amplitude
0.94 relative to the fiducial model. The fiducial model is well within
the 1$\,\sigma$ (68\%) confidence region (the difference in $\chi^2$
between the fiducial model and the best-fit model is 0.64 for 2
df).  The actual data and the best fit model are shown in
Fig.~\ref{fig:phase}C. This test shows that our data are entirely
consistent with the model predictions, and that we can rule out (at
$\approx 3\,\sigma$) a pathological model in which the $EE$
oscillations are in phase with $TT$ rather than out of phase.

An alternate, and more physically motivated, way to look at the phase
of the peaks in $EE$ is to use fits to the fiducial model spectrum of the form
\begin{equation}
l(l+1)C_l^{EE} = (A_s/A_{s0}) \,  \left(f(\ell \theta /\theta_0) + g(\ell \theta /\theta_0) \sin(k\ell \theta /\theta_0)\right) .    
\end{equation}
This parameterizes the models in terms of two of the cosmological
parameters discussed earlier, $A_s$ and $\theta$. The values of these parameters in the fiducial model are $A_{s0}$ and
$\theta_0$ ($\theta_0=1.046$). Changing $\theta$ scales the whole function,
including the envelope, rather than just the phase.  We now examine
the variation of $\chi^2$ as these two parameters are changed, the
other  four cosmological parameters being fixed at their fiducial
values.  There is a minimum of $\chi^2$ near the fiducial model, with
$\theta/\theta_0 =1.02\pm0.04$ and $A_s/A_{s0}=0.93$. (A
second minimum in which the third polarization peak is shifted and
scaled to fit the second fiducial peak is incompatible with
the $TT$ data.) This test also shows that the $EE$ data strongly prefer
the fiducial model, and demonstrates that the $EE$ data alone have the
power to place constraints on cosmological parameters.

\paragraph*{Tests for systematics.}
      We have carried out a number of data quality tests to look for
      possible systematic contamination by foreground
      emission, residual ground emission, or other instrumental
      effects.  We have found no evidence of significant residual
      instrumental or foreground effects after correcting for the
      point sources and projecting out the common ground spillover
      mode.

\begin{figure}
\includegraphics[width=\textwidth]{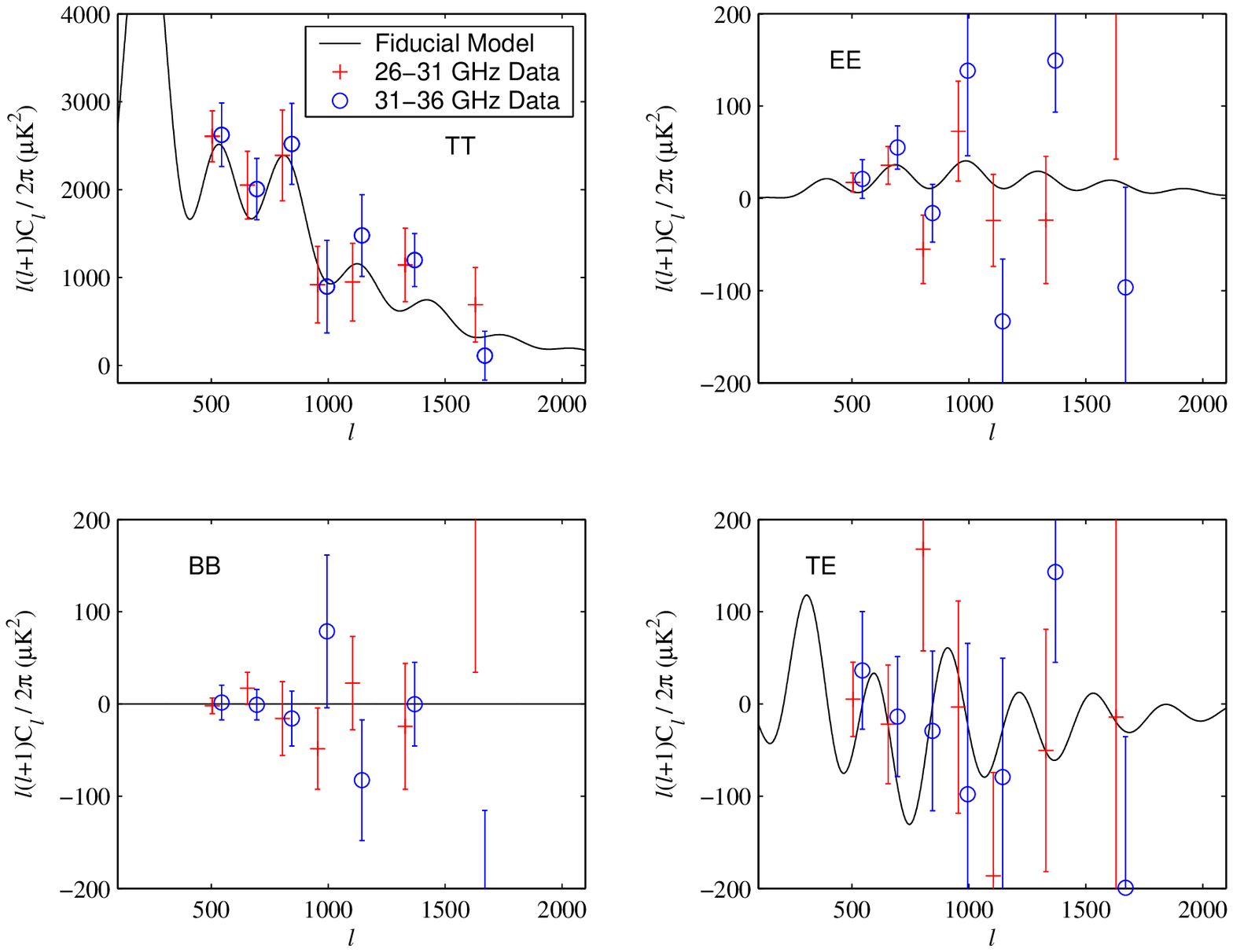}\\
\caption{CBI polarization power spectra obtained from low- and high-frequency
  channels. {\it Red points}: 26--31 GHz; {\it blue points}: 31--36
  GHz. The points have been offset horizontally in $l$ for clarity.  The
  four panels show  total intensity power spectrum $TT$,
  curl-free polarization mode power spectrum $EE$, curl
  polarization mode power spectrum $BB$, and cross-spectrum $TE$.  The
  fiducial model curve is the same as in Fig.~6.}
\label{fig:chanspectra}
\end{figure}

      Foreground emission is likely to have a different
      spectrum from the CMB, and ground contamination is
      frequency-dependent because it depends strongly on the baseline
      length in wavelengths, and thus shows up most on the shortest
      baselines at the lowest frequency.  To look for these effects we
      estimated power spectra separately from the data taken in the
      lower and upper halves of our frequency band, i.e., 26 to 31 and
      31 to 36 GHz (fig.~\ref{fig:chanspectra}).  We have compared the two spectra
      by using $\chi^2$ (including the bin-to-bin correlations).  For
      $EE$ and $BB$ the measurements are dominated by thermal noise
      (rather than sample variance) so the $\chi^2$ results are valid.
      The results are:  $\chi^2 = 8.43$ (7 df) for $EE$, and
      $\chi^2 = 8.30$ (7 df) for $BB$. The
      probability of obtaining a larger $\chi^2$ by chance is 0.30 for
      $EE$ and 0.31 for $BB$. The power spectra thus show no indication 
      of strong contamination by foreground emission or residual
      ground emission. Note that for $TT$ and $TE$ the maximum likelihood
      error estimates include the effects of sample variance, and, because
      sample variance is correlated between the two frequency bands, a
      simple $\chi^2$ test is not valid. 
      
      In addition to dividing the data into two frequency bands, we
      carried out jackknife tests in which we compared the
      following subsets of the data.  (i) We
      compared all subsets of three of the four fields. This would
      indicate whether any of the fields is anomalous and is a good
      test for foreground contamination. No significant differences
      were found, and in particular the $14^{\rm h}$ field (which lies
in the North Polar Spur region) was not
      anomalous. (ii) We compared all subsets of 12 of the 13 antennae.
      This would show up problems associated with particular antennae
      or receivers.  (iii) We compared the $TT$ spectra derived from the
      $R$ and $L$ antennae separately, to check for calibration
      discrepancies.  (iv) We compared spectra estimated from the first
      and second halves of the data set, to check for effects based on
      season, distance from the primary calibrator, and other
      time-dependent parameters. None of these tests showed any
      significant differences between the data subsets.
      
      The DASI results increase our confidence that diffuse
      synchrotron emission is not a significant contaminant in our
      $EE$ spectrum. The fields that we have observed appear to be
      comparable to the DASI fields (Fig.~\ref{fig:galaxy}), and the DASI
      95\% confidence upper bound of $0.91\,\mu$K$^2$ on $EE$ contamination
      should also apply to the CBI observations, which were made at higher
      $l$ where the contribution of synchrotron emission is expected to
      be lower.


\section*{Conclusions}

\begin{figure}
\includegraphics[width=\textwidth]{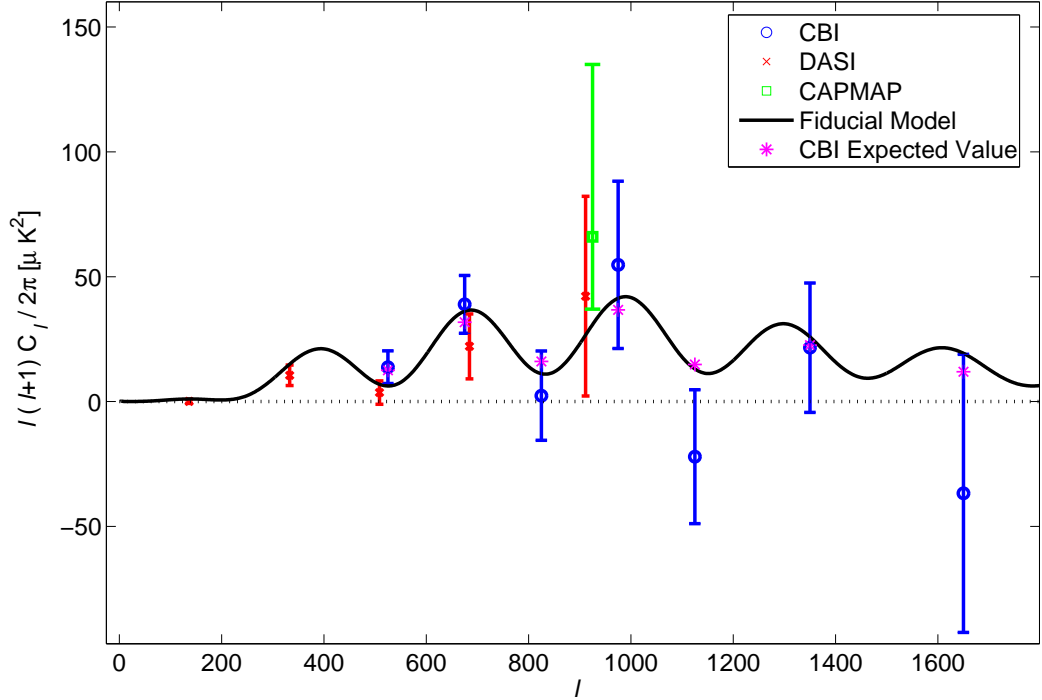}
\caption{Comparison of $EE$ measurements from CBI, DASI
  \cite{Leitch04}, and CAPMAP \cite{Barkats04}. The fiducial model curve
  is the same as in Fig.~\ref{fig:spectra}. The asterisks show the
  predictions of the fiducial model for the CBI bands.}
\label{fig:comparison}
\end{figure}

Our $EE$ results are shown in comparison with the recent results from
DASI and CAPMAP in Fig.~\ref{fig:comparison}.  We have detected the
polarized CMB ($EE$) emission with high confidence ($8.9\,\sigma$ when
foreground sources are ignored and $7.0\,\sigma$ when potentially
contaminating sources are projected out), and we have also measured
the phase of the $EE$ spectrum and shown that it is consistent with a
phase-shift of $\pi$ relative to the $TT$, as expected if acoustic
waves are the origin of the features in the $TT$ and $EE$ spectra on
the scales of clusters of galaxies.  The results from the CBI and DASI
experiments are a powerful confirmation of the predictions of the
standard model. 
The CBI continues to observe the polarized CMB emission, and we expect
by the end of 2005 to have more than doubled the data set, leading to
a decrease of over a factor two in the uncertainties of $C_l$.

\section*{Supporting Material}

\paragraph*{Modifications for polarization measurement}


The polarization observations reported in this paper were made between
September 2002 and May 2004. Earlier tests of the technique were made
using a single cross-polarized antenna
\cite{Cartwright_thesis,Cartwright04}.
The CBI was upgraded in 2002 to enhance its polarization capability.
This involved replacement of the existing polarizers with new
broadband achromatic polarizers, replacement of the
high-electron-mobility-transistor (HEMT) amplifiers with new
lower-noise amplifiers, and reconfiguration of the antennae into a
more compact array.

The circular polarization mode ($R$ or $L$) received by each antenna
can be selected by changing the orientation of a quarter-wave plate in
front of the low noise amplifier.  The original CBI quarter-wave
plates were replaced by achromatic DASI-style polarizers
\cite{Kovac_thesis,Leitch02} that could be rotated under computer
control so that the polarization in any antenna can be changed in
$<5$~s. An important design goal for polarization observations is to
limit the polarization impurity: if an antenna does not receive pure
$R$ or $L$ polarization the linear polarization measurements will be
corrupted by an admixture of total intensity $I$. The fraction of $I$
that appears in the $RL$ or $LR$ visibility measurement is called the
leakage. This is a complex number (amplitude and phase) that
must be measured for each antenna.  The new polarizers reduced the
leakages from $\sim 5$--$15\%$ to $\sim 1$--$3\%$.  The leakages are
stable, and typically exhibit changes of $<0.2\%$ over periods of a
few months.

\begin{figure}
\includegraphics[width=\textwidth]{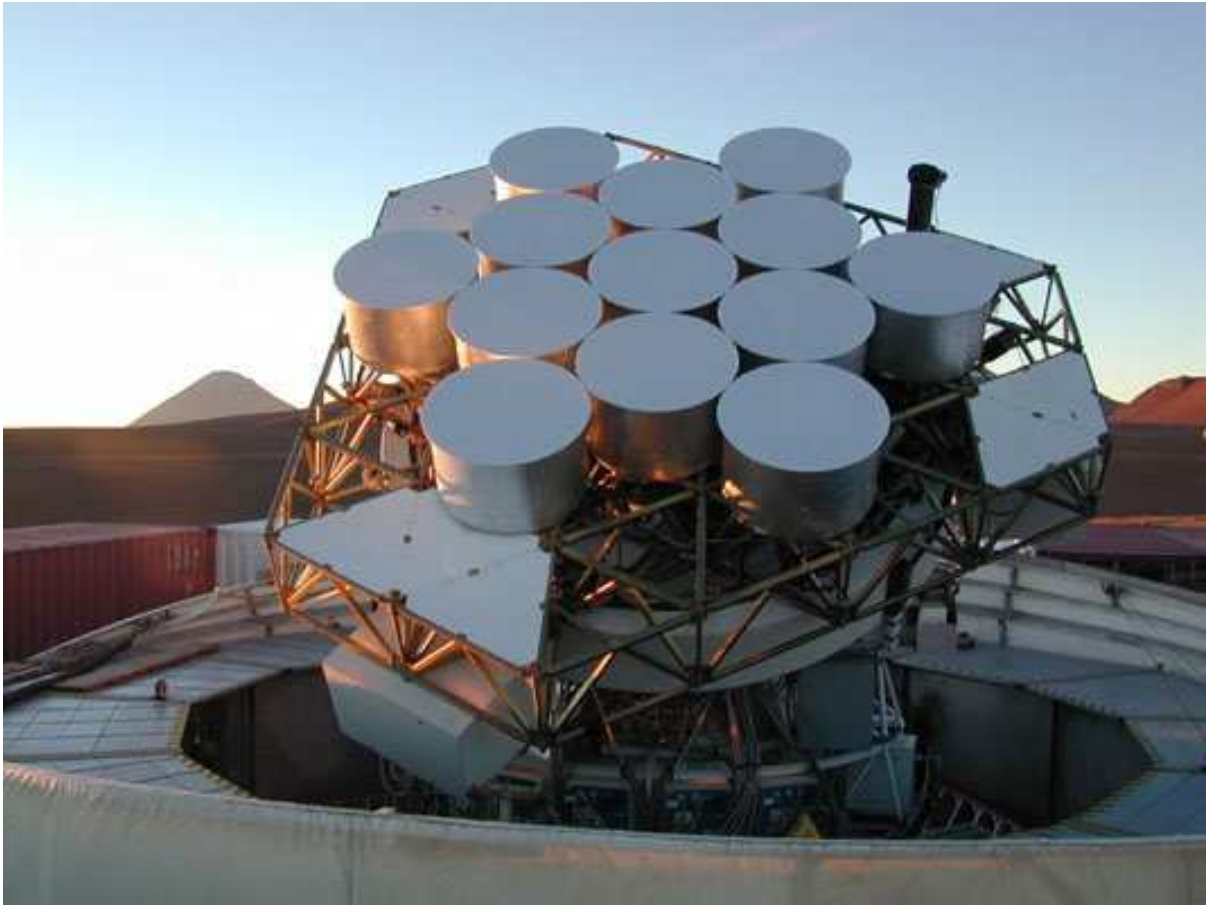}
\caption{The Cosmic Background Imager. For the polarization observations,
  the 13 90-cm Cassegrain antennae were arranged in this hexagonal
  close-packed configuration on the rotating, alt-az mounted platform,
  with six adjusted to be sensitive to right-hand circularly polarized
  radiation and seven to left-hand circularly polarized radiation. By
  correlating the signals from the antennae in pairs, 78
  interferometer baselines are obtained ranging in length from 1.0 to
  3.5~m.}
\label{fig:cbi}
\end{figure}

In the first two years of operation of the CBI (2000 and 2001) we used
sparse configurations of the antennae in order to cover a wide range
of multipoles ($300<l<3500$).  For the polarization observations we
decided to concentrate on the multipole range $300$--$2000$ in order
to provide maximum sensitivity in the region where the CMB polarized
$EE$ signal is expected to be the greatest.  For this reason we
adopted the close-packed configuration shown in Fig.~\ref{fig:cbi}
This configuration provides the highest concentration of short
baselines possible with the CBI and provides an excellent match to the
$l$ range of the expected maximum $EE$ signal.



\paragraph*{Data calibration and editing.}

During the observations and initial data analysis, we inspected each
night's observations to look for instrumental and other problems. For the
final analysis, we used automatic procedures to remove data with known
problems (warm or unstable receivers, for example) and with higher than
normal noise levels. This last check eliminated $\approx 1\%$ of the
data that had been corrupted by clouds or instrumental problems.

\paragraph*{Amplitude calibration.}

For the CBI, the amplitude and phase calibration of the co-polar
visibility data ($RR$ or $LL$) was carried out using the same
procedures as for the 2000--2001 observations \cite{Mason03}.  The
refinement of the CBI flux density scale using the WMAP observations
of Jupiter has been described in \cite{Readhead04}.  The uncertainty
in the revised scale is 1.3\% in flux density (2.6\% in the power
spectrum, $C_l$).  On most nights one or more of the primary
calibration sources Tau~A (the Crab Nebula), Jupiter, Saturn, and
3C\,274 was observed. All of these gave consistent results, except for
3C\,274: we found that the flux density of 3C\,274 declined by $ 7\%$ over
the period of these observations. The model for 3C\,274 was adjusted to
take this secular variation into account before the final calibration
of the data.

The majority of CBI data are calibrated using measurements of Tau~A and
Jupiter, as in \cite{Readhead04}.
When none of the primary calibration sources was available we used
secondary calibration sources, such as the variable quasar J1924$-$293,
for which we obtained flux densities by interpolating from adjacent
days calibrated against the primary calibrators.

\paragraph*{Polarization calibration.}

The calibration measurements on the co-polar baselines yield
complex gain factors for each antenna.  These gain factors are
sufficient to calibrate the cross-polarized baselines ($LR$ and $RL$)
except for an unknown phase difference between the $R$ antennae and
the $L$ antennae, equivalent to an unknown rotation of the plane of
linear polarization \cite{Leitch02,Cotton99}.  
We determine the unknown $L-R$
phase difference by observations of a strong, polarized calibration
source, Tau~A, for which we assume the polarization position angle
($E$-vector) is $-27.6^\circ$.  This value was derived
\cite{Cartwright_thesis,Cartwright04} by comparison of CBI and Very Large
Array \cite{VLA} observations of 3C\,273 and 3C\,279, both of which vary
but are observed regularly with the CBI at 26--36 GHz and the VLA at
frequencies straddling the CBI band (22 GHz and 44 GHz)
\cite{vlapolarweb}. It is close to the angle measured with other
instruments at lower frequencies. The $L-R$ phase difference is very
stable unless receivers or cables are modified, so on nights when no
measurement of Tau~A was available we used the average of all the
Tau~A measurements.

\paragraph*{Leakage measurement}

We measured the instrumental polarization leakage factors on each
night when either Tau~A or Jupiter could be observed. These
observations were made at a number of different parallactic angles, by
rotating the CBI platform, to enable the source and instrumental
polarization to be separated. The instrumental leakage was found to be
in the range 1\%--3\% on most baseline--channels, with a few
baseline--channels showing leakages as high as 5\%. We determined that
leakage did not vary significantly across the field of view by making
observations of Tau~A at a number of offset positions.  This low level
of instrumental polarization and our strategy of rotating the deck so
that many antenna pairs contribute to the same $(u,v)$ point ensures
that any instrumental polarization in our final data set is
negligible ($<1\%$).  We have therefore ignored the instrumental polarization
in the present analysis.

\paragraph*{Noise calculation.}

In order to ensure that the ground contamination was identical in each
of a set of six pointings, we deleted all visibility samples that did
not have counterparts observed at the same hour angle (within the
tolerance of the integration time, 4.2~s) in all of the six
fields. After selecting matched data points in this way, we calculated
the noise from the scatter of the visibility measurements. As an error in the
noise estimate will bias the final power spectrum estimate, it is important to obtain an accurate estimate of the noise in the data.
In one
scan, comprising observations of 6 fields, we record $m=1,\ldots,M$
($M$ varies, but is usually about 37) data points (complex
visibilities) for each of $n=1,\dots,N$ fields ($N=6$) . The observed
visibility $V_{nm}$ is related to the true visibility $X_{n}$ and the
ground contribution $g_m$ by
\begin{equation}
V_{nm} = X_n + g_m + r_{nm},
\label{eq:vis}
\end{equation}
where $r_{nm}$ is the noise in the measurement. Note that $X_n$ is the
same for all $m$ (we do not change the baseline length or orientation
relative to the sky during the scan), and $g_m$ is the same for all
$n$ (the ground contribution is assumed to be the same in each field,
i.e., constant for the duration of the scan at a given elevation and
azimuth). An estimator for the noise variance is
\begin{equation}
{1\over(M-1)(N-1)} \sum_{n=1}^N \sum_{m=1}^M \Psi_{nm}^2.
\end{equation}
where $\Psi_{nm}$ is derived from $V_{nm}$ by subtracting the mean of
the $N$ measurements from all $N$ obtained at each time $m$, and the
mean of the $M$ measurements from all $M$ obtained on each field
$n$. Our best estimator is the average of the two estimates obtained
by treating the real and imaginary parts of the visibility separately.
We obtained a single noise estimate for each baseline--channel
that applies to a whole 18~min scan.  The variance of the estimator is
\begin{equation}
\VAR{\widehat{\sigma^2}} = {\sigma^2 \over (M-1)(N-1)}.
\label{eq:varvar}
\end{equation}
The uncertainty in the noise estimate in each scan is small enough
that noise bias \cite{Mason03} is not a concern in the present
observations. Scans with rms noise more than three times that expected
for normal system temperatures were deleted; in most cases the high
noise was due to clouds.

\paragraph*{Power spectrum estimation}

The principles of estimating polarization power spectra from
interferometer visibility measurements are described by
\cite{Kovac02}. To process the CBI data, we have extended the
gridding-based procedure used in our earlier work \cite{Myers03} to
deal with mosaicked polarization observations.  A given correlator output
sample, or visibility, can be one of the four polarization products
$RR$, $RL$, $LR$, or $LL$.  These can be related to the fundamental
CMB polarization modes $T$ (temperature), $E$, and $B$ (polarization)
through the expressions given in Equations 3 and 4 of \cite{Kovac02}.
The resulting power spectra are decomposed into the six possible
covariances $TT$, $EE$, $BB$, $TE$, $TB$, and $EB$.  Note that because
the CBI measures circular polarization products, which are orientation
independent (depending only on the handedness of the
wave polarization), the CBI (or any interferometer using circularly
polarized receptors) is sensitive to the $E$ and $B$ modes directly.
This simplifies the power spectrum analysis.

 The co-polar $RR$ and $LL$ visibilities are gridded together into an effective
$RR$ estimator ($\langle LL\rangle$ and $\langle RR\rangle$ are identical in
the absence of circular polarization) as in \cite{Myers03}, while the
cross-polar $RL$ and $LR$ visibilities are gridded together, after conjugating
and reflecting the $LR$ visibilities in the $uv$-plane, into cross-polar
estimators $\bvec{\Delta}_{RL}$ using the same gridding kernel as the co-polar
data.  The covariance matrix elements are computed for the cross-polar
estimators using a modified operator
${\bf P}_{RL}(\bvec{v}) ={\bf P}(\bvec{v})\,e^{i\,2\,(\chi-\psi)}$ 
where ${\bf P}$ is defined in Equation~12 of \cite{Myers03},
$\psi$ is the on-sky parallactic angle  of the CBI receivers
(Equation~2 of \cite{Leitch02}) and $\chi$ is the wave-vector angle
corresponding to the $uv$ point $\bvec{v}$ (Equation~3 of \cite{Kovac02}).
The band powers derived from the likelihood analysis are then $\{ q^{\rm S}_B,
B =1,\ldots,N^{\rm S}_B \}$; the different covariance products 
$S=TT,EE,BB,TE,TB,EB$ can have different numbers and locations of bands.  Point
sources are handled in the same manner as in \cite{Myers03}, with the option
of projecting out the sources from the $RL$ and $LR$ parts of the covariance.
The new scanning procedure required the addition of a scan projection matrix
${\bf C}^{\rm scan}$ constructed by building a ``noiselike'' matrix 
as in Equations~32 and 35 of \cite{Myers03}, with
the covariance elements $E_{kk^\prime}=1$ if visibilities $k$ and $k^\prime$
are from the same scan (otherwise zero); this is then projected out with
a pre-factor $q_{\rm scan}$ in the same way as the point sources are.
Details of this modified procedure will be
given in \cite{Myers04}.

The maximum likelihood estimation of the spectrum from the gridded
estimators is done on the CITA McKenzie cluster \cite{Dubinski03}
which consists of 256 nodes with two 2.4 GHz Intel Xeon processors and
1 GB of memory per node.  The matrix operations are done using the
SCALAPACK library \cite{slug}.  From an initial guess of the spectrum,
we iterate to the maximum likelihood solution using the Newton-Raphson
method.  One modification to the procedure used in \cite{bjk98}
provides a significant improvement.  Rather than use the standard
approximation to the curvature, with which the number of expensive
matrix operations is proportional to the number of bins in the spectrum,
we use an approximate
curvature that requires only a single matrix inversion
\cite{Sievers_thesis}.  Using 32 nodes per mosaic, with $10^4$
estimators per mosaic, this decreases the time per iteration from
about an hour to one minute, without changing the solutions.  The
total time for the spectrum to converge, once the estimators and
correlations are read into memory, is about 10 minutes, and is
virtually independent of the number of bins in the spectrum.

\clearpage

\begin{table}
\caption{CBI Band Powers in $\mu$K$^2$. The broad first band is not sensitive over the 0--600 range, but with
        finer binning we have a detection in a band near 400.}
\label{tbl:powers}
\scriptsize
\begin{verbatim}
                  TT                  EE                  BB                  TE
   l-range       Power     Error     Power     Error     Power     Error     Power     Error

     0   600    2882.0     276.8      13.7       6.5      -0.7       5.5       8.7      32.1
   600   750    1865.8     275.8      38.9      11.6       7.2       8.1     -15.6      41.0
   750   900    2413.4     342.7       2.4      17.9      -1.1      16.7      17.9      57.6
   900  1050    1098.0     306.5      54.8      33.5      14.6      30.8     -96.7      76.6
  1050  1200    1411.4     303.6     -22.1      26.8      20.8      30.7     -57.1      67.1
  1200  1500     988.9     189.1      21.6      25.9      23.6      26.9     -49.6      54.2
  1500  5000     191.4     164.2     -36.8      55.7     -55.5      51.8      29.4      91.4


\end{verbatim}
\end{table}

\begin{table}
\caption{Cosmological constraints from the WMAP1 only, CBI pol plus
   WMAP1, and CBI pol plus CBIext plus WMAP1 data compilations for an assumed
   $\Omega_{tot}=1.0$. Relaxation of this constraint opens up the tight
   uncertainties on $H_0$ and $\Omega_m$.
   We included weak external priors on the Hubble
   parameter ($45 \, {\rm km\, s^{-1} \, Mpc^{-1}} <H_0<90$ ${\rm km\,
   s^{-1} \, Mpc^{-1}}$) and the age of the universe ($t_0 > 10 \, {\rm
   Gyear}$). The flatness prior has the strongest effect on the
   parameters by breaking the geometrical degeneracy and allowing us to
   derive tight constraints on $H_0$ and $\Omega_m$. The top six
   parameters are those used in the Markov chain evaluations and the
   distributions of the bottom six are derived from the same chains. The
   uncertainties are given as 68\% confidence intervals.}
\label{tbl:paramtab}
\vspace{1em}
\begin{center}
\begin{tabular}{c|ccc}
\hline\\
& WMAP1 & CBIpol+WMAP1 & CBIpol+CBIext+WMAP1\\
\\
\hline\\
 $        \Omega_b h^2$ & $ 0.0243^{+0.0019}_{-0.0017} $& $ 0.0240^{+0.0018}_{-0.0016} $ & $ 0.0233^{+0.0013}_{-0.0013} $        \\
 $        \Omega_c h^2$ & $ 0.119^{+0.016}_{-0.016} $   & $ 0.113^{+0.014}_{-0.015} $    & $ 0.109^{+0.012}_{-0.013} $        \\
 $          \theta$ & $ 1.049^{+0.007}_{-0.008} $   & $ 1.048^{+0.006}_{-0.006} $    & $ 1.044^{+0.005}_{-0.005} $        \\
 $            \tau_c$ & $ 0.188^{+0.037}_{-0.065} $   & $ 0.190^{+0.044}_{-0.067} $    & $ 0.164^{+0.027}_{-0.053} $        \\
 $             n_s$ & $ 1.01^{+0.06}_{-0.05} $      & $ 1.00^{+0.06}_{-0.05} $       & $ 0.98^{+0.04}_{-0.04} $        \\
 $\log[10^{10}A_S]$ & $  3.3^{+0.2}_{-0.2} $      & $  3.3^{+ 0.2}_{- 0.2} $       & $  3.2^{+ 0.2}_{- 0.2} $        \\
 $  \Omega_\Lambda$ & $ 0.72^{+0.08}_{-0.07} $      & $ 0.74^{+0.07}_{-0.07} $       & $ 0.75^{+0.06}_{-0.06} $        \\
              Age (Gyr) & $ 13.3^{+ 0.4}_{- 0.4} $      & $ 13.4^{+ 0.3}_{- 0.4} $       & $ 13.5^{+ 0.3}_{- 0.3} $        \\
 $        \Omega_m$ & $ 0.28^{+0.07}_{-0.08} $      & $ 0.26^{+0.07}_{-0.07} $       & $ 0.25^{+0.06}_{-0.06} $        \\
 $        \sigma_8$ & $ 0.94^{+0.13}_{-0.13} $      & $ 0.91^{+0.10}_{-0.10} $       & $ 0.85^{+0.08}_{-0.08} $        \\
 $          z_{re}$ & $ 17.5^{+6.7}_{-6.2} $      & $ 17.5^{+ 6.7}_{- 6.2} $       & $ 16.0^{+ 6.0}_{- 5.5} $        \\
 $             H_0$ & $ 73.3^{+7.1}_{-6.4} $      & $ 74.5^{+ 7.7}_{- 6.5} $       & $ 74.2^{+ 6.1}_{- 5.5} $        \\
\\
\hline
\end{tabular}
\end{center}
\end{table}

\clearpage

\begin{thebibliography}{99}
\bibitem{multipole}
{The power spectrum of a function defined on the celestial sphere is computed
  by spherical harmonic analysis, in which the multipole number $l$ in
  analogous to the wavenumber in Fourier analysis. The dipole has $l=1$, and
  higher $l$ correspond to smaller angular scales, $\theta \approx 180^\circ/l$
  \cite{Hu02}.}
\bibitem{note_peaks}
{The first peak has very low amplitude and has not yet been detected.}
\bibitem{cluster}
{Our typical run involves the calculation of 16 separate Markov chains. Each
  chain is run on a separate  node of the CITA McKenzie Beowulf cluster
  for a typical run-time of approximately 3 hours. The chains are run until the
  largest eigenvalue returned by the Gelman-Rubin convergence test reaches
  0.01.}
\end{thebibliography}
\end{document}